# Temperature-dependent Hysteresis in MAPbI$_3$ Solar Cells

Igal Levine[1], Pabitra K. Nayak[2], Jacob Tse-Wei Wang[2], Nobuya Sakai[2], Stephan Van Reenen[2], Thomas M. Brenner[1], Sabyasachi Mukhopadhyay[1], Henry J. Snaith,[2] Gary Hodes[1]* David Cahen[1]*

[1]Dept. of Materials & Interfaces, Weizmann Inst. of Science, Rehovot, Israel, 76100
[2]Clarendon Laboratory, Dept. of Physics, Univ. of Oxford, Oxford, UK, OX1 3PU

**TOC**

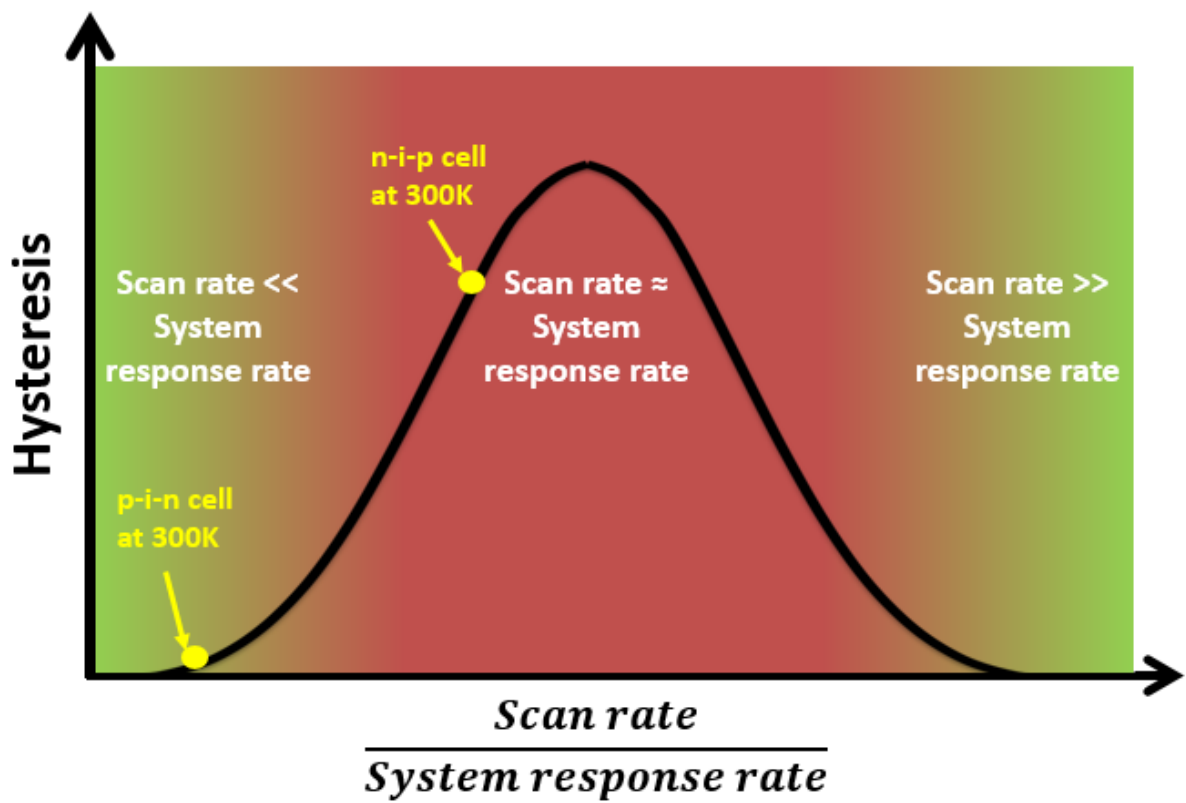

---

* * Corresponding author email: david.cahen@weizmann.ac.il

**Abstract** Hysteresis in the current-voltage characteristics of hybrid organic-inorganic perovskite-based solar cells is one of the fundamental aspects of these cells that we do not understand well. One possible cause, suggested for the hysteresis, is polarization of the perovskite layer under applied voltage and illumination bias, due to ion migration *within the perovskite*. To study this problem systemically current-voltage characteristics of both regular (light incident through the electron conducting contact) and so-called inverted (light incident through the hole conducting contact) perovskite cells were studied at different temperatures and scan rates. We explain our results by assuming that the effects of scan rate and temperature on hysteresis are strongly correlated to ion migration within the device, with the rate-determining step being ion migration at/across the interfaces of the perovskite layer with the contact materials. By correlating between the scan rate with the measurement temperature we show that the inverted and regular cells operate in different hysteresis regimes, with *different* activation energies of 0.28±0.04 eV and 0.59±0.09 eV, respectively. We suggest that the differences, observed between the two architectures are due to different rates of ion migration close to the interfaces, and conclude that the diffusion coefficient of migrating ions in the inverted cells is 3 orders of magnitude higher than in the regular cells, leading to different accumulation rates of ions near the interfaces. Analysis of $V_{OC}$ as a function of temperature shows that the main recombination mechanism is trap-assisted (Shockley-Read Hall, SRH) in the space charge region, similar to what is the case for other thin film inorganic solar cells.

## Introduction

Light to electrical power conversion efficiencies of hybrid organic-inorganic lead-halide perovskite-based solar cells have grown rapidly over the past few years, starting from ~3% in 2009 to > 20% recently.[1] The exceptional efficiencies of MAPbI$_3$–based cells have been attributed to many factors[2,3], among them the high degree of crystallinity of these solution-processed absorbers, their very high extinction coefficient and sharp optical absorption onset, low defect state density within the band gap and long diffusion length[4]. A shadow over the MAPbI$_3$ based solar cells is their stability, especially the often seen hysteresis in their photovoltaic current-voltage characteristics. Overcoming this problem requires understanding of the mechanism(s) that give rise to hysteresis. Among possible origins, suggested for hysteresis, we note:

(1) ferroelectric polarization of the perovskite layer.[5,6] This is questionable, because as yet no direct proof for ferroelectric behavior of Halide Perovskites (HaPs), including MAPbI$_3$, has been found under the conditions that the materials are used in PV cells[7];

(2) charge accumulation at the interfaces due to trapping and de-trapping [8,9,10,11,12] and

(3) ion migration of MA$^+$/ I$^-$ - related defects under applied bias,[13,14,15] which, naturally, could also be a cause for (2).

Here we focus on the hysteresis observed in two types of MAPbI$_3$ solar cells, to help understand the origin(s) by performing J-V scans at different temperatures, scan rates, and illumination intensities and analyzing the results with the help of modeling. The devices were so-called 'n-i-p' and 'p-i-n' ones, i.e., with compact TiO$_2$ and Spiro-OMeTAD as the n- and p-type contact layers, and with PEDOT:PSS (poly(3,4-ethylenedioxythiophene) : polystyrene sulfonate) and PCBM (phenyl-C61-butyric acid methyl ester) as the p and n-type contact layers, respectively, based on whether the contact through which the cell is illuminated is the 'n' or the 'p' one. Our study demonstrates how supposedly "hysteresis-free" p-i-n devices actually do show hysteresis in the J-V curves, depending on the temperature and scan parameters, and we conclude that interface-dependent ionic migration/accumulation, is the main cause for the observed hysteresis in MAPbI$_3$ cells. By correlating between the scan rate and the temperature we also show that the rate of ion migration at the interfaces of the perovskite layer is the rate-determining step and is highly dependent on the -p or -n contact materials. Furthermore by analyzing the different time-scales in which hysteresis is observed in the n-i-p and p-i-n cells, we show that two different diffusion coefficients for ion migration are obtained, $10^{-12}$ and $10^{-9}$ cm$^2$ s$^{-1}$ for the n-i-p and p-i-n cells, respectively, suggesting that although the different cell structures contain the same absorber material, the *different contact materials* determine the rate of ion migration.

**Experimental section**

**Sample preparation**

FTO-coated glass sheets (7Ω/sheet, Pilkington) were used as transparent electrode and substrate. To obtain the required electrode pattern, FTO layer was etched with zinc powder and 2M HCl. The sheets were then washed with 2% Hellmanex in water, deionized water, acetone, ethanol and iso-propanol, respectively followed by cleaning in oxygen plasma for 10 min.

p-i-n device: Precursor solutions for CH$_3$NH$_3$PbI$_3$ were prepared separately by dissolving lead iodide (PbI$_2$) in DMF (450 mg ml$^{-1}$), and MAI in isopropanol (50 mg ml$^{-1}$), respectively. The hole transporting layer (~20 nm) was deposited from PEDOT:PSS (AI 4083, HERAEUS) solution on to the patterned FTO sheets. The CH$_3$NH$_3$PbI$_3$ perovskite layer was prepared following methods described elsewhere.[16] To form CH$_3$NH$_3$PbI$_3$ on a PEDOT:PSS layer, a PbI$_2$ layer was first deposited by spin-coating at 6000 rpm for 30s from its precursor solution, followed by drying at 70°C for 5 min. Then a MAI layer was deposited on the dried PbI$_2$ layer by spin coating at 6000 rpm for 30 s from a precursor solution, followed by annealing in a N$_2$ glove box for 1 hour. The n-type collection layer, PCBM, dissolved in dichlorobenzene, 2 wt% was spin-coated on top of the perovskite layer, followed by spin-coating a buffer layer of bathocuproine (BCP) from its solution in isopropyl alcohol. Finally, devices were completed by thermal evaporation of silver (70 nm) as electrical contact.

n-i-p device: The patterned FTO sheets were coated with a TiO$_2$ compact layer (~80 nm) by spin-coating using a titanium sol precursor at 2000 rpm for 1 min. The titanium sol precursor was prepared by dropwise adding a solution of 350 µL of tetra-titaniumisopropoxide in 2.5 mL of ethanol into a solution of 35 µL of 2M HCl solution in 2.5 mL of ethanol. The coated FTO substrate was then heated to 300 °C for 20 min, then ramped up to 500 °C for 20 min. 38 wt% perovskite precursor solutions were made by dissolving MAI and $PbCl_2$ (in a 3:1 molar ratio) in dimethylformamide (DMF). Perovskite precursor solution was coated onto the FTO substrate with a compact $TiO_2$ layer by a two-step spin-coating process i.e., at 1400 rpm and 3000 rpm for 20 and 15 s, respectively, under controlled 15-17 % relative humidity. The resultant thin film was allowed to dry at room temperature for 15 min then at 70 °C for 10 min. The dried perovskite film was annealed at 100 °C for 90 min and then at 120 °C for 15 min in a box oven. Spiro-OMeTAD (2,2(7,7(-tetrakis-(N,N-di-pmethoxyphenylamine)9,9(-spirobifluorene))) was used as hole transporting material. 98 mg of spiro-OMeTAD was dissolved in 1 ml chlorobenzene along with 10 µl of tert-butylpyridine and 32 µl of lithium bis(trifluoromethylsulfonyl)imide salt solution (170 mg ml$^{-1}$) in acetonitrile. An Ag metal contact layer was deposited as the top electrode on the HTM layer by thermal evaporation.

**Sample characterization**

The electrical measurements were carried out in a cryogenic probe station (TTPX Lakeshore). First the sample was placed at room temperature (RT) in a vacuum of ~2x10$^{-4}$ mbar, and then the sample was cooled down to 200K, and kept at that temperature for 1 h for stabilization. J-V curves were recorded by a Keithley 6430 sub-fA SourceMeter unit. The probe station was equipped with micromanipulator probes using BeCu tips with a diameter of 3 µm. J-V data were recorded at 200 K, 240 K, 280 K and 320 K; at each temperature the sample was allowed to stabilize for 1h prior to recording the J-V scans. Since studies have shown that the J-V curve of these cells can be greatly influenced by the conditions prior to the measurement (voltage/light-soaking),[17] we wanted to avoid such pre-biasing effects, and performed all the measurements under the same conditions, regardless of the order of the measurements. Therefore all the J-V measurements were done in the following order – starting from higher to the low light intensity, in steps of 0.01 V for the 5, 50 and 300 mV sec$^{-1}$ scan rates and 0.1 V for the 1500 mV sec$^{-1}$ scan rate. At each temperature, starting from the high to low light intensities, the sample was first allowed to reach equilibrium by illumination for 60 s under open-circuit conditions; then the slowest scan was performed (5 mV sec$^{-1}$), followed by another 60 s under illumination at open-circuit, then a faster scan was performed, and this procedure was followed for each of the light intensities.

During the temperature-dependent measurements of the p-i-n devices the vacuum varied between 1x10$^{-5}$ to 1x10$^{-4}$ mbar. Illumination was by a white light LED (no UV or IR emission). Illumination intensities were equivalent to 0.60, 0.39 and 0.14 sun, i.e., lower than one sun equivalent, and were calibrated using a Si solar cell. The illuminated area was 0.13 cm$^2$ for both types of the studied solar cells.

**Results**

**Figure 1** shows the comparison between the TiO$_2$ / MAPbI$_3$ / spiro-OMeTAD, 'n-i-p' device, and the PEDOT:PSS / MAPbI$_3$ / PCBM, 'p-i-n' device, at RT.

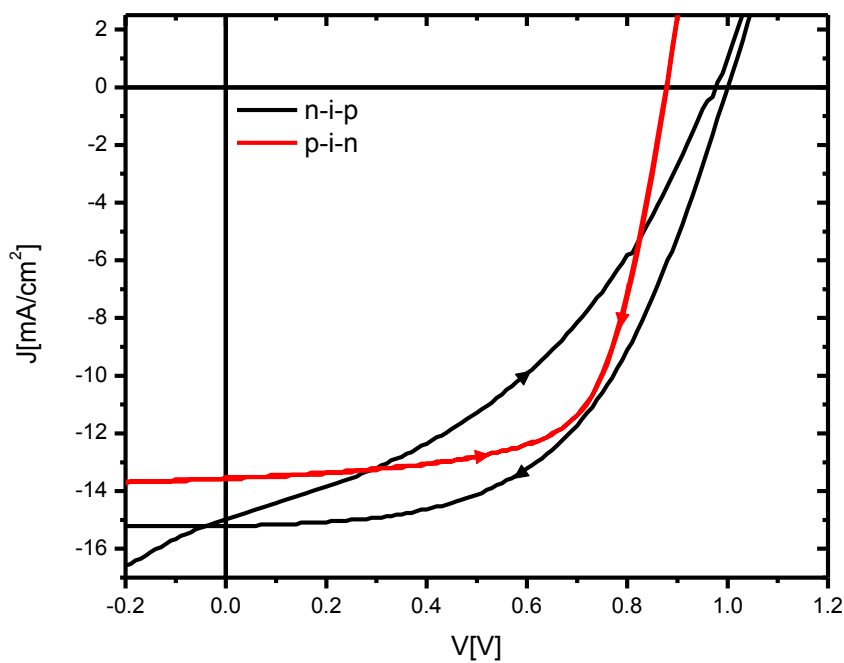

*Figure 1: J-V curves of the studied p-i-n cell (red) and n-i-p cell (black), measured under 60 mW cm$^{-2}$ light intensity at a scan rate of 50 mV s$^{-1}$ at room temperature. Arrows indicate the scan directions.*

As can be seen from Figure 1, there is essentially no hysteresis at RT for the p-i-n device in this study, in agreement with previous reports,[18,19] which showed that in this architecture, no hysteresis occurs at RT, while for the n-i-p devices hysteresis is observed mainly in the *FF* and slightly in the $V_{OC}$ and $J_{SC}$.

### 1. Temperature dependence of the p-i-n cells

Figure 2 shows the J-V characteristics of the p-i-n cells in different temperatures, at a constant scan rate of 50 mV s$^{-1}$ and 0.6 sun illumination.

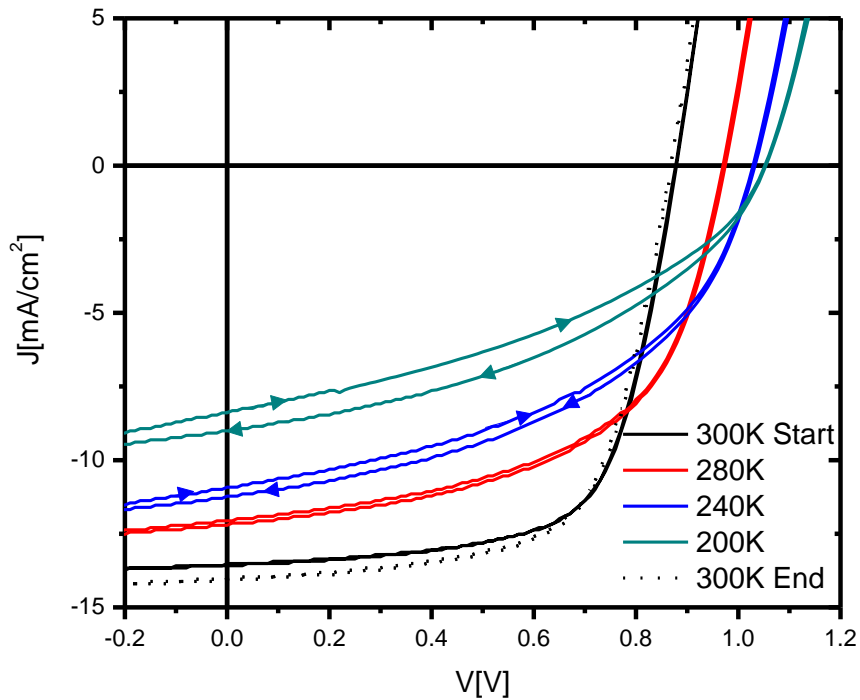

*Figure 2: J-V curves of p-i-n device measured under light intensity of 60 mW cm$^{-2}$ at a scan rate of 50 mV s$^{-1}$; dashed line represents the J-V curve taken after heating back up to 300K; arrows indicate the scan directions.*

The results, shown in Figure 2 indicate that as the temperature is lowered, several changes occur:

(1) $V_{OC}$ increases, as expected from theory, and discussed further below.

(2) The fill factor and $J_{SC}$ decrease, probably because the conductivities of the p- and n-type contact materials decrease with decreasing temperature, increasing the series resistance. The temperature dependence of charge transport within the MAPbI$_3$ layer needs also to be considered, but because at low temperature both the carrier mobility and the lifetime of the perovskite increase [20] this cannot explain the decrease in conductivity with decreasing temperature;

(3) No irreversible degradation is observed, and the cell performance returned to its original performance after the 300 to 200 to 300 K temperature cycle (dashed black line).

(4) Hysteresis starts to appear, becoming more pronounced at low T, but disappears when the temperature is raised back to 300 K.

We thus studied the effects of the temperature *and* scan rate *and* illumination intensity on the extent of hysteresis observed in the J-V curves under illumination for the p-i-n devices and results are shown in Figure 3, which shows the J-V data measured at 200 K and 320 K (for the J-V data between 240-280 K and intermediate light intensity, 39 mW cm$^{-2}$, see the SI), for the studied p-i-n device.

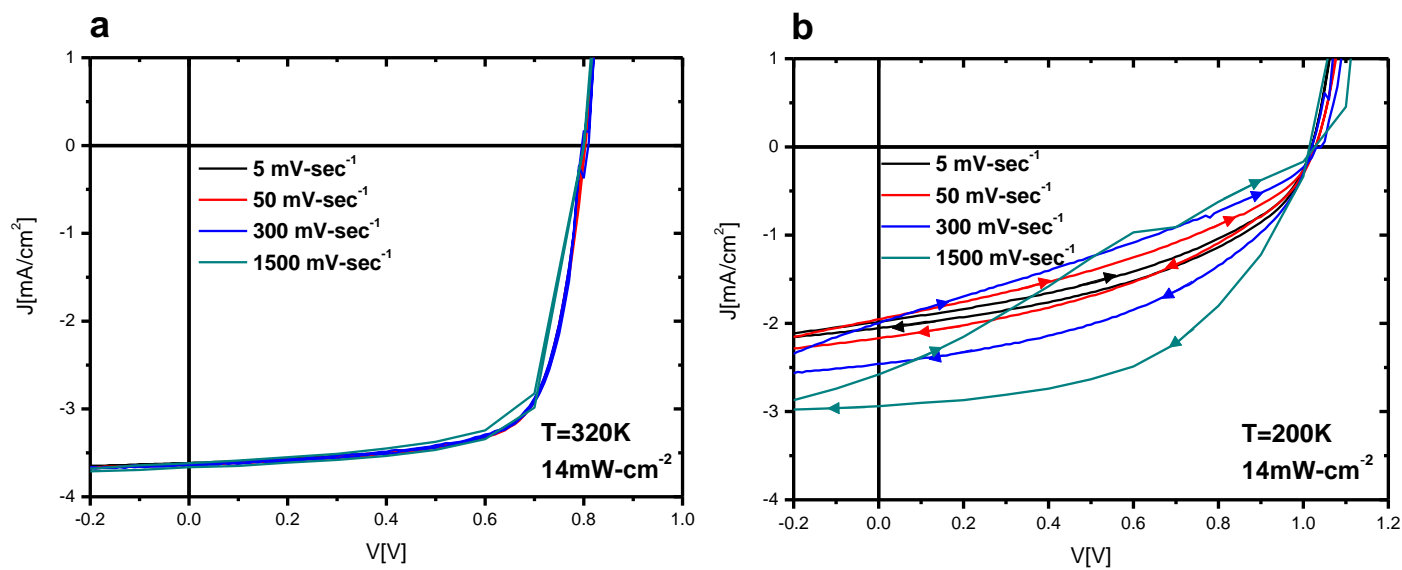

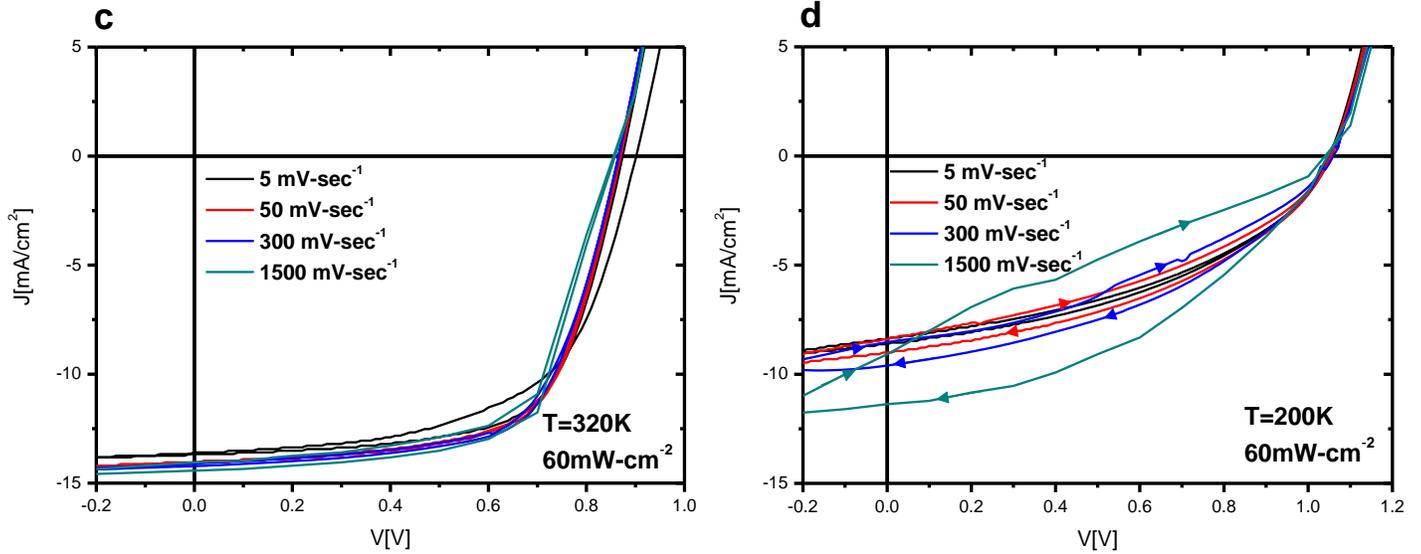

*Figure 3: J-V curves of p-i-n device measured under low and higher light intensity (**a, b**: 14 mW cm$^{-2}$ ;**c, d**: 60 mW cm$^{-2}$) and at different scan rates at 320K (**a, c**) and 200K (**b, d**); arrows in **b** and **d** indicate the scan directions.*

In order to find a correlation between the hysteresis and scan rate/temperature, we chose to quantify the hysteresis observed in Figure 3 in each of the curves in a comparable manner by defining an empirical hysteresis factor (*HF*) as follows:

$$HF \equiv \left| 1 - \frac{A_{Jsc \to Voc}}{A_{Voc \to Jsc}} \right| \tag{1}$$

where *A* denotes the area under the *J-V* curve for either the scan from $J_{SC}$ to the $V_{OC}$ (backward scan) or the scan from the $V_{OC}$ to the $J_{SC}$ (forward scan). We quantified the hysteresis in such a way so as to reflect contributions from Voc AND Jsc AND the FF, instead of various existing definitions in the literature[12,21] which use the current density at a specific voltage point.

Figure 4 illustrates the behavior of *HF* as a function of the scan rate, light intensity and temperature.

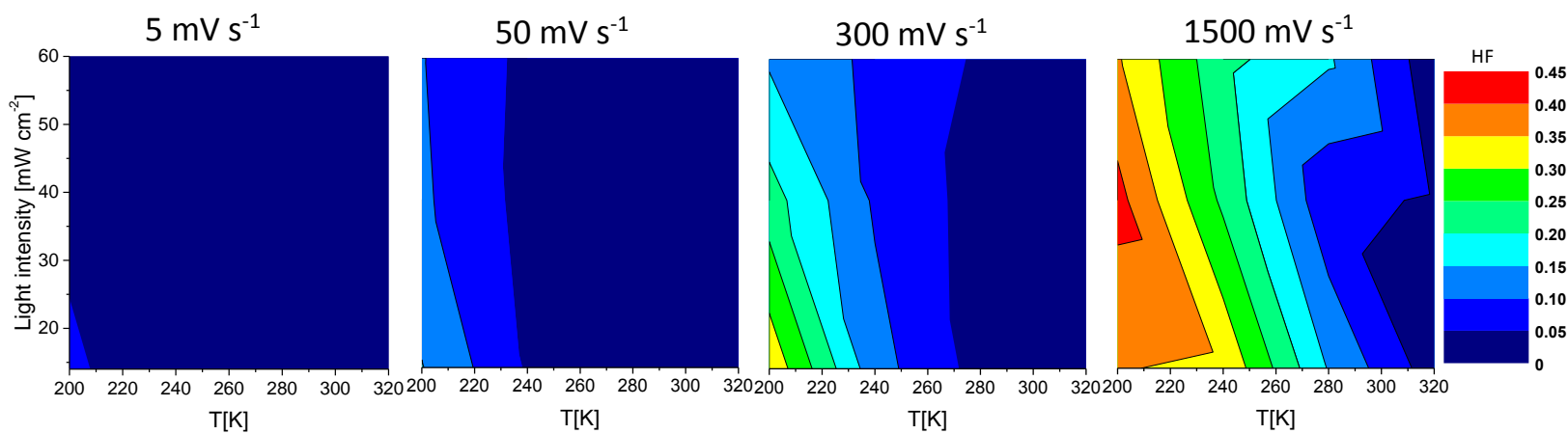

*Figure 4: Hysteresis factor (HF, as defined is Equation (1), as a function of light intensity and temperature for scan rates of 5 (a) 50 (b) 300 (c) and 1500 mV s$^{-1}$ (d). The color map indicates the value of the hysteresis factor ranging from 0 (dark blue, no hysteresis) to 0.45 (red, maximum hysteresis) for the p-i-n device.*

The main trends observed in Figure 4, are:

1. The hysteresis factor increases with increasing scan rate.
2. The hysteresis factor is higher for lower temperatures.

### 2. Temperature dependence of the n-i-p cells

Unlike the case with the p-i-n devices, when n-i-p devices are measured from 300 K down to 240 K, their performance not only degrades upon lowering the temperature, showing a pronounced increase in series resistance, but also does not return to the initial state upon heating back to RT (Figure S5). This indicates that an irreversible change occurred to one or more of the layers/interfaces when measuring these cells between 300 and 240 K in vacuum, and this hinders a systematic study of hysteresis in such devices as function of temperature over this temperature range. Furthermore, when n-i-p cells are measured in vacuum without cooling (only slight heating to 320 K), a pronounced degradation is observed (Fig. S6). These results suggest that measuring the n-i-p cells in vacuum might de-dope the spiro-OMeTAD layer by removal of oxygen[22] and, hence significantly, lower its conductance and increase the series resistance (an explanation that requires further investigation, beyond the scope of this work). Thus, to study the effect of temperature on the n-i-p cells, measurements were performed in a $N_2$ environment, over a limited temperature range, between 280 K (lower limit prior to ice formation on the sample) and 330 K (upper limit, due to the phase change from tetragonal to cubic structure). If measurements were done in ambient air for prolonged duration ( > 15 min) at slightly elevated temperatures, substantial degradation was observed, as shown in Figure S7).

Figure 5a shows the *J-V* curves of the n-i-p device, between 280 K to 330 K, under 0.6 sunlight intensity in an $N_2$ environment, with 50 mVs$^{-1}$ scan rate. The effect of the scan rate at different temperatures is shown in Figures 5b-d.

The following trends are observed:

(1) $V_{OC}$ increases with decreasing temperature, as for p-i-n cells, as expected from the general diode model and discussed further below.

(2) Hysteresis is largest at the lowest temperature, 280 K, and decreases with increasing temperature.

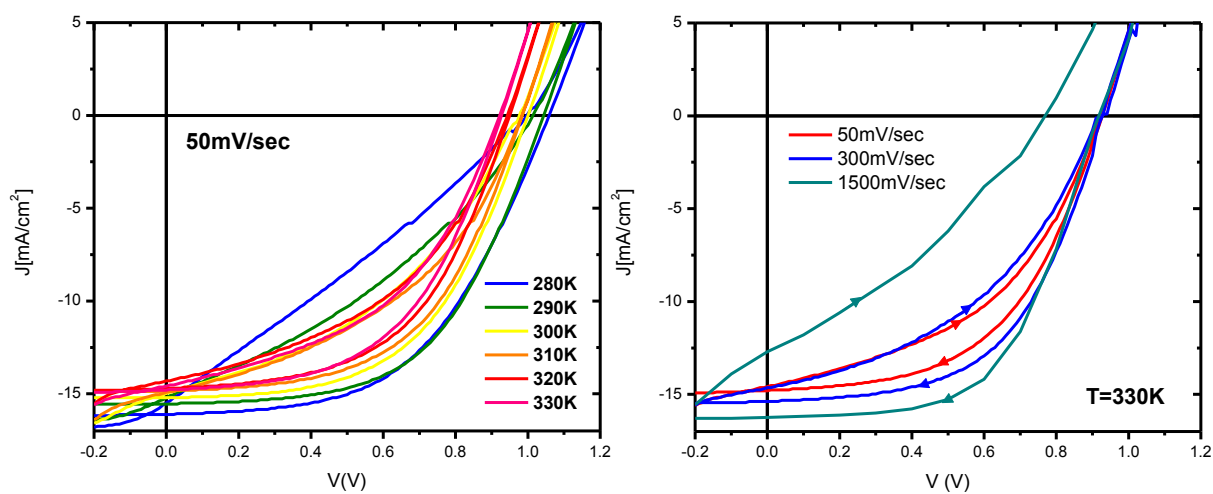

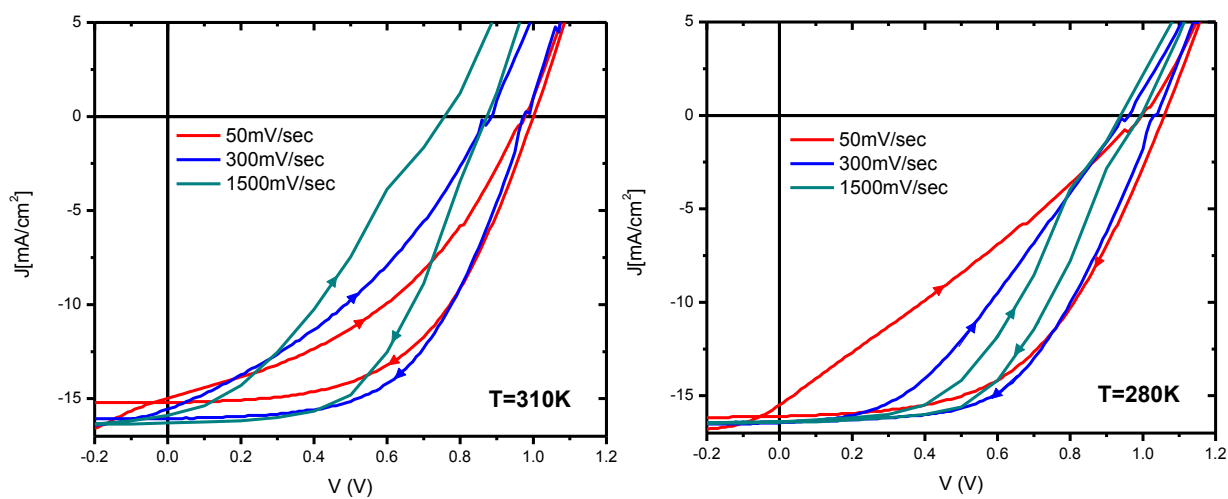

*Figure 5: J-V curves of n-i-p device measured under 60 mW cm$^{-2}$ light intensity at different temperatures at a scan rate of 50 mV s$^{-1}$ (**a**) and at different scan rates at 330 K (**b**), 310 K (**c**) and 280 K (**d**); arrows in **b, c** and **d** indicate the scan directions.*

As can be seen from Figure 5b-d, the effect of scan rate and temperature in n-i-p cells is more complex than in p-i-n ones, and in order to better observe the interplay between the scan rate and temperature, the corresponding HF maps, as shown in Fig. 4 for the p-i-n cells, are plotted in Figure 6.

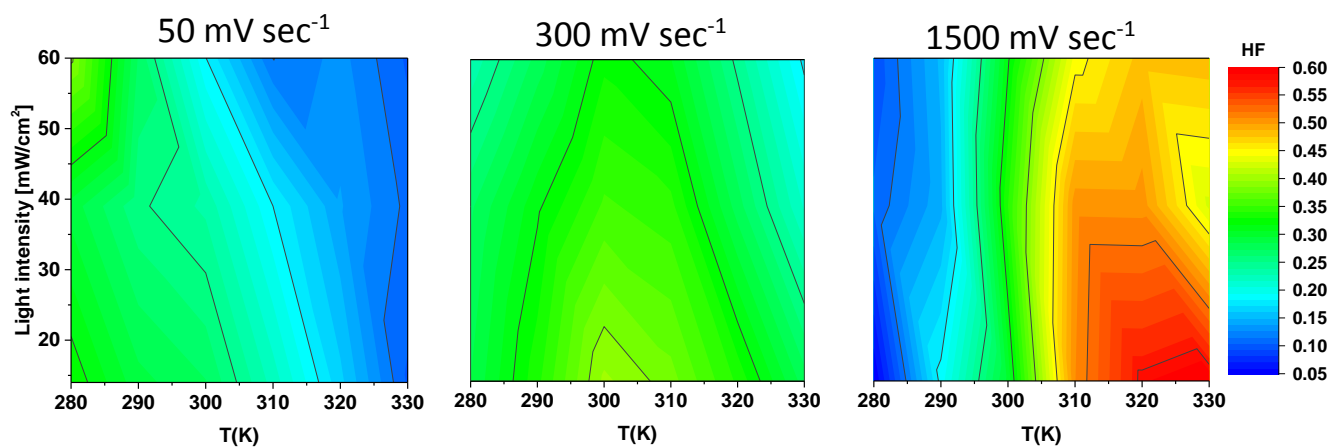

*Figure 6: Hysteresis factor (HF, as defined is Equation (1)) for n-i-p devices as a function of light intensity and temperature for scan rates of 50 (a) 300 (b) and 1500 mV s$^{-1}$ (c). The color map indicates the value of the hysteresis factor ranging from 0.05 (least hysteresis) to 0.60 (red, maximum hysteresis). See also Fig. S8 in the Suppl. info. for more detailed 2-D plots for each of the scan rates.*

Although the temperature range used to study the n-i-p cells (60 K) is narrower than that used for the p-i-n ones (120 K), it can already be seen from figure 6 that the temperature dependence of the HF is stronger for the n-i-p cells, which suggests that in this case the temperature-activated process, which is the cause for the hysteresis, has a <u>*significantly higher*</u> activation energy, $E_A$, than in the p-i-n case. We also note that there is little dependence of the HF on the light intensity in the range of scan rates and temperature that we could access and we cannot exclude that stronger effects may occur if the light intensity is varied over several orders of magnitude.

### 3. Modelling results

A numerical model was used to simulate the effect of temperature on the J-V hysteresis in halide perovskite solar cells. While temperature will affect ion mobility in a rather straightforward manner (i.e., increase of ion mobility, $\mu$, with increasing temperature, $\mu \propto e^{\frac{-E_A}{kT}}$), changes in the extent of hysteresis due to temperature-dependent trapping is a more complicated process, and we thus used numerical modelling to help identify such effects.

Basically, the model describes generation, recombination, transport, trapping, injection and extraction of electronic and ionic carriers in a full perovskite solar cell. More details on this model can be found in the Supporting information. Simulations were performed similarly to those reported in Ref.[23] where it was shown that hysteresis originates from ion redistribution, that depends on the bias history of the cell. The ion distribution affects the spatial distribution of free carriers by its effects on the electrostatic potential landscape[24,25]. The redistribution of free carriers leads to changes in the trapped carrier densities near the perovskite interface(s). The non-radiative recombination rate at recombination centres near the perovskite interface(s) is proportional to the trapped carrier densities of both signs. Thus, overall, the ion distribution affects the non-radiative recombination rate near the perovskite interface(s), which results in the appearance of hysteresis in the J-V curve.

To model temperature effects due only to traps, (ion transport was not modelled due to excessively long calculation times) on hysteresis in perovskite solar cells, devices were first allowed to stabilize (i.e., reach "steady state") at either $V_{stabilized}$ = 0.0 V or 1.2 V. During this time, ions were

*Submitted to Advanced Energy Materials, November 1st, 2015*

allowed to move by setting their mobility at $10^{-1}$ cm$^2$V$^{-1}$s$^{-1}$. This high value has no further effect on the calculations and was used to speed up calculations (and is still orders of magnitude lower than the slowest electronic process). The ion transport was allowed to reach steady-state, indicated by the ion current density approaching zero, thus simulating the case of a very slow (<< 1mV s$^{-1}$) J-V sweep. Note that after steady-state conditions are reached, the ion distribution is independent of ion mobility as ion drift and diffusion, both of which are proportional to the ion mobility, are then in equilibrium. Furthermore, the ion densities only vary by maximally 10 % for the temperature range used here, independent of ion mobility, indicating no significant difference in ion density for the different temperatures. After reaching steady-state, the ion mobility was set to zero and a homogeneous electronic charge carrier generation rate ($G$) was applied to the perovskite layer, resulting in approximately 1 sun generation rate in total. Then the bias voltage was swept to record *I-V* characteristics, which are plotted in Figure 7a. These characteristics were calculated at 5 different temperatures, i.e., $T$ (K) = 200, 250, 300, 325, 350 for $V_{stabilized}$ (V) = 0.0 and 1.2. In these simulations, no temperature effects related to ion motion are possible as the ions are fixed. Therefore these results only show the temperature effects related to other processes.

The results show clear hysteresis effects that depend on the temperature. At higher temperatures, the hysteresis loops are observed to shrink and $V_{oc}$ is also reduced.

The dependence on temperature can be rationalized by a temperature-dependent detrapping rate of trapped carriers at the interface between the perovskite and the electron extraction layer, taken to be proportional to exp(-q$E_t$/$kT$), where $E_t$ is the interfacial trap energy (which is interpreted as a theoretical activation energy). Figure 7a shows the temperature dependence of the hysteresis when the trap depth is set at 0.2 eV, demonstrating that the hysteresis increases as the temperature decreases due to a decreased de-trapping rate. Modelling the hysteresis for various trap depths in the perovskite (Figure 7b) shows that hysteresis is enhanced by deeper trap depths. The results in Figure 7a and b, taken together, show that the largest hysteresis effect is obtained when the detrapping rate (~exp(-$E_t$/$kT$)) is as low as possible, either by reducing the temperature or increasing the trapping energy, $E_t$. We note that the large temperature effect shown in Figure 7a would not be present for larger trap energies as the detrapping rate will rapidly become negligible with increasing $E_t$, as shown in Figure 7c. Already for a 0.1 eV larger trap depth, i.e., $E_t$ = 0.3 eV, the detrapping rate decreases ~ 60 x at 300K (exp(-0.1/$kT$)). Thus, our model suggests weaker temperature dependence of the HF as the trap depth increases.

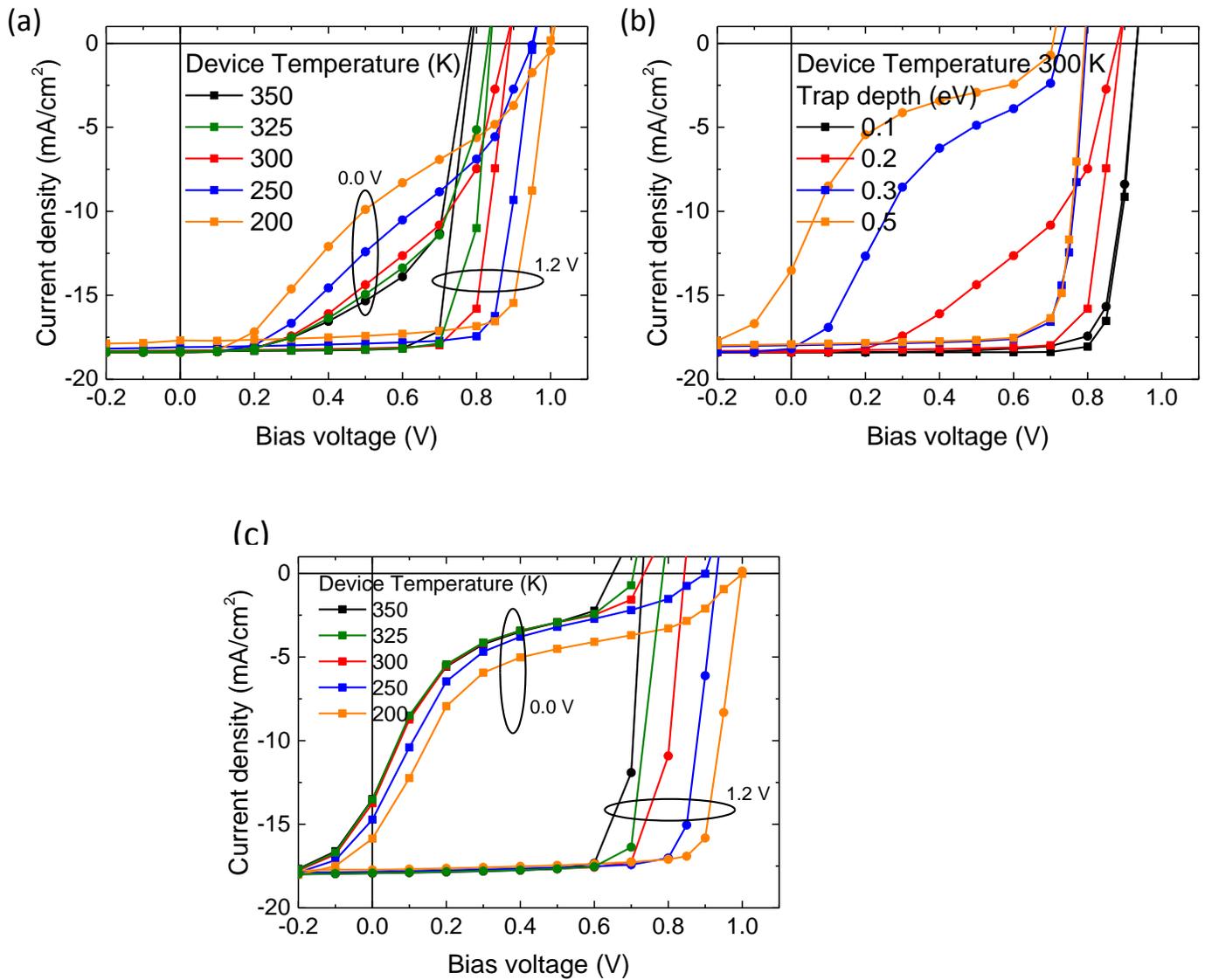

*Figure 7: J-V characteristics of modelled halide perovskite solar cells at G = 2.5 nm⁻³s⁻¹ for **a**) varying temperature and 0.2 eV trap depth and **b**) different trap depth energies at 300 K. **c**) varying temperature and 0.5 eV trap depth; The value of V$_{stabilized}$ is indicated for the corresponding characteristics in **(a,c)**.*

**Discussion**

The experimental results in Figures 3-6 demonstrating the interplay between hysteresis, scan rate and temperature, both for the n-i-p and p-i-n devices, show that hysteresis in the J-V curve for a device containing MAPbI$_3$ is highly dependent on the J-V scan conditions. Tress et al. showed that the hysteresis in n-i-p devices (comprising TiO$_2$ and spiro-OMeTAD as the n and p contact layers, respectively), increases with increasing scan rate at RT,[17] as observed in Figure 3 for the p-i-n device, and in Figure 5c for the n-i-p device. They concluded that the main reason for this phenomenon is a so-called "slow" process, probably ion migration, which does not follow the change in applied voltage when high scan rates are employed. Further suggestive evidence for ion migration was provided by Beilsten-Edmands et.al[7], Yang et al.[14] and Eames et al.[13], who claimed that iodide vacancies are mobile in the perovskite and can move in an electric field. This ion movement could lead to build-up of ions near the contacts under the influence of an electric field (by the internal built-in field or by an external applied bias). It has been well-established by Riess et. al. in a series of papers over the past decades that show how such ion build-up will alter the electric field within the device.[26–28] Depending on the direction of the change, this can improve the charge collection efficiency when scanning from $V_{OC}$ to $J_{SC}$ compared to the scan from $J_{SC}$ to $V_{OC}$, as seen from Figures 3 and 5 and agrees well with previous findings.[13] The result is hysteresis in the *J-V* curves, which is mainly expressed via the *FF* and the $J_{SC}$. Recently Brynt et al.[21] showed that by measuring p-i-n cells at low temperatures, one can also see the emergence of hysteresis in the *J-V* curves upon cooling, as we also observe in this study. Charge trapping at one of the MAPbI$_3$ interfaces was also suggested to cause hysteresis in the J-V curves via non-radiative trap-assisted recombination[29,30], and the expected effect of the trap depth (energy) according to our model is shown in Figure 7b. The temperature dependence of the trap population will lead to enhanced hysteresis for reduced temperature.

While for the p-i-n device, hysteresis clearly increases with increasing scan rate, or reduced T, in the n-i-p device the results shown in Figures 5-6 suggest that there are two ways to decrease hysteresis in the J-V curves: (1) scanning at a low scan rate (≤ 50 mV s$^{-1}$) at high T, or (2) scanning at very high scan rates (≥ 1500 mV s$^{-1}$) at low T. This observation suggests that under the conditions of (1), the slow process, which is accelerated due to the elevated temperature, can follow the relatively low scan rate better than at lower T, leading to less hysteresis, while under the conditions of (2), the slow process is decelerated due to the lower temperature, and thus at high scan rates almost cannot follow at all, leading again to less hysteresis. These considerations lead to the conclusion that there should be some intermediate scan rate/temperature conditions for which maximum hysteresis is observed. Therefore our experimental results, in which the effects of scan rate and trapping-de-trapping kinetics of charge carriers are coupled to one another and cannot be separated, show more hysteretic effects than those predicted by the simulations (which exclude ion motion during the J-V scan).

This rationalization suggests plotting HF as a function of only *one* variable that contains both the scan rate and the temperature. Such a variable, *R*, can be defined empirically and to a first order approximation, as the ratio between the scan rate and the response rate of the system (system response rate):

$$R \equiv \frac{Scan\ rate}{System\ respone\ rate}$$

A schematic illustration of the expected behavior of HF as a function of *R* is shown in Figure 8:

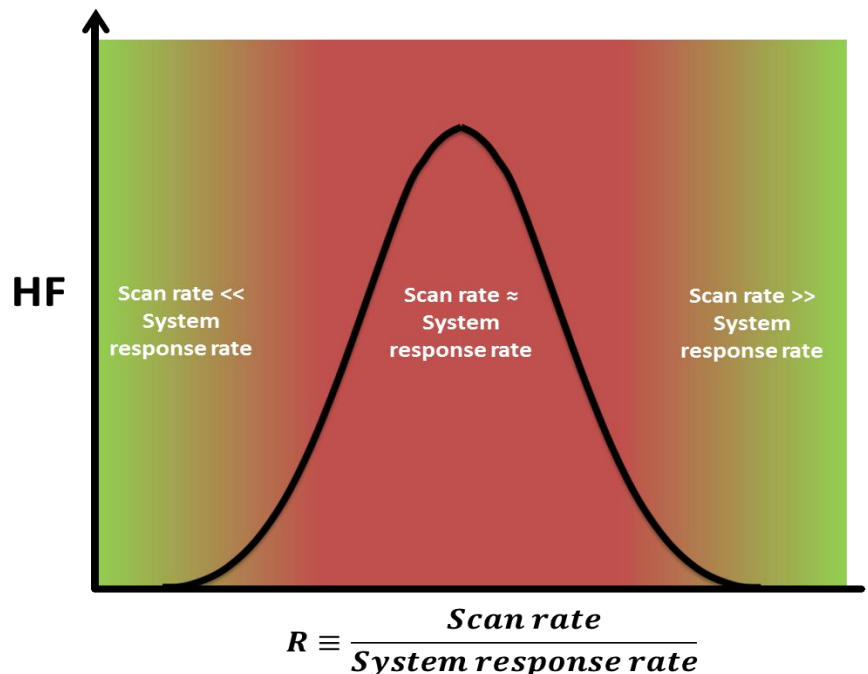

*Figure 8: Schematic illustration of the HF (for either p-i-n or n-i-p cells) as a function of R, defined as the ratio between the scan rate to the response rate of the system*

Each point on the X axis of Figure 8 represents both a scan rate *and* a temperature, which will determine the extent of the observed hysteresis in the J-V curve (HF value). Qualitatively, one can see from the HF maps of the p-i-n (Figure 4) and n-i-p (Figure 6) devices that in the n-i-p device, all the 3 regimes depicted in Figure 8 are observed experimentally, while for the p-i-n device, only the left and center parts are observed. In order to quantitatively plot *HF*(*R*), an expression is needed for the slow process rate, and as a first order approximation, one can take the rate of the rate-limiting step (slowest process occurring) to have an Arrhenius-like form: $k \propto e^{-\frac{E_A}{kT}}$, where $E_A$ is the activation energy for the system response rate. To estimate the activation energy, which, based on the different temperature dependences of the p-i-n and n-i-p devices, differs between them, one can use the scan rate as an effective rate constant, which defines the rate constant "imposed" on the system.

Since in the *J-V* scan we are measuring the current at every voltage point, we can start observing hysteresis when the response time of the "slow" process, whatever its nature, does not keep up with the voltage scan rate. In order to monitor the extent of the hysteresis in a systematic manner we changed (experimentally) both the scan rate and the temperature in such a way that a time lag would occur between the voltage sweep and the response of the sample, i.e. by measuring at a very high scan rate (1500 mV s$^{-1}$ in our case), during which the system does not reach voltage-current steady state, or by slowing the slow process by lowering the temperature. The fact that at a scan rate of 5 mV s$^{-1}$ we do not see hysteresis in the studied temperature/light intensity range for the p-i-n case, suggests that at this scan rate, the system reaches a steady state for every measured point, even at 200 K, and the time to reach this "steady state" can be estimated to be between 10 mV / (5 mV s$^{-1}$ ) = 2s (typical voltage step, lower limit) to 3000 mV / (5 mV s$^{-1}$ ) = 10 minutes (entire voltage range, upper limit). This demonstrates that the so-called "slow process" is occurring on a timescale of a few seconds to minutes, much slower than electronic conduction and recombination processes (<μs), as observed by others.[17,31,32,33]. Because ionic migration is highly temperature dependent, our results are consistent with the hypothesis (cf., for example, [13,14,34]) that such migration is necessary for the observed hysteresis in MAPbI$_3$ cells, and is the rate-limiting step in the process. At higher scan rates, when the dwell time at each voltage (which we calculate by dividing the voltage step – 10 mV, by the scan rate) is less, i.e., 200, 33 and 6.7 ms for the 50, 300 and 1500 mV s$^{-1}$ scans, respectively, it is less likely that the system will reach steady-state before taking the next current measurement at the next voltage point. This rationalization allows us to estimate an activation energy required to reach the steady-state for a given scan rate in the p-i-n device. If we take the scan rate and consider it an "effective rate constant" by which the system reaches steady-state, we would expect that at every rate constant (scan rate), there will be a different temperature onset at which hysteresis will begin to appear (*HF* increasing from 0 to 0.05), which can be seen as the transitions from dark blue to blue in Figure 4 for each of the scan rates. We see that this transition is reached at higher temperatures as the scan rate increases, reaching nearly 310 K at the highest scan rate of 1500 mV s$^{-1}$ (roughly similar for all light intensities ±10 K). Thus, by plotting the logarithm of the scan rate (effective rate constant) vs. 1/k*T*, where *T* is the observed temperature for the transition to *HF* = 0.05, we can experimentally extract an activation energy, $E_A$, of the slow process for the p-i-n devices. Note that we estimate the error for the transition temperature to *HF* = 0.05 from Figure 4, to be ± 10 K, which takes into account the deviation from an abrupt transition between the regions).

Because the same exercise cannot be performed for the n-i-p device, as the HF was larger than 0.05 at all the different scan rates studied, instead, the common value of HF for the 3 scan rates used can be the maximum point of the observed hysteresis at each scan rate at the maximum light intensity used, 60 mW cm$^{-2}$. Then for the n-i-p device, the slope of a plot of the logarithm of the scan rate vs. (kT)$^{-1}$, where T is the temperature at which HF reaches the maximum for each scan rate in Figure 6a-c, (using the contour lines in Figure S8) we estimate the maxima to be at 280 K, 300 K and 325 K, for 50, 300 and 1500 mV s$^{-1}$, respectively, with an estimated error of ±5 K) will yield the corresponding $E_A$. Figure 9 shows the resulting activation energies for both types of devices, yielding $E_A$ = 0.28±0.04 eV and 0.59±0.09 eV for the p-i-n and n-i-p devices, respectively.

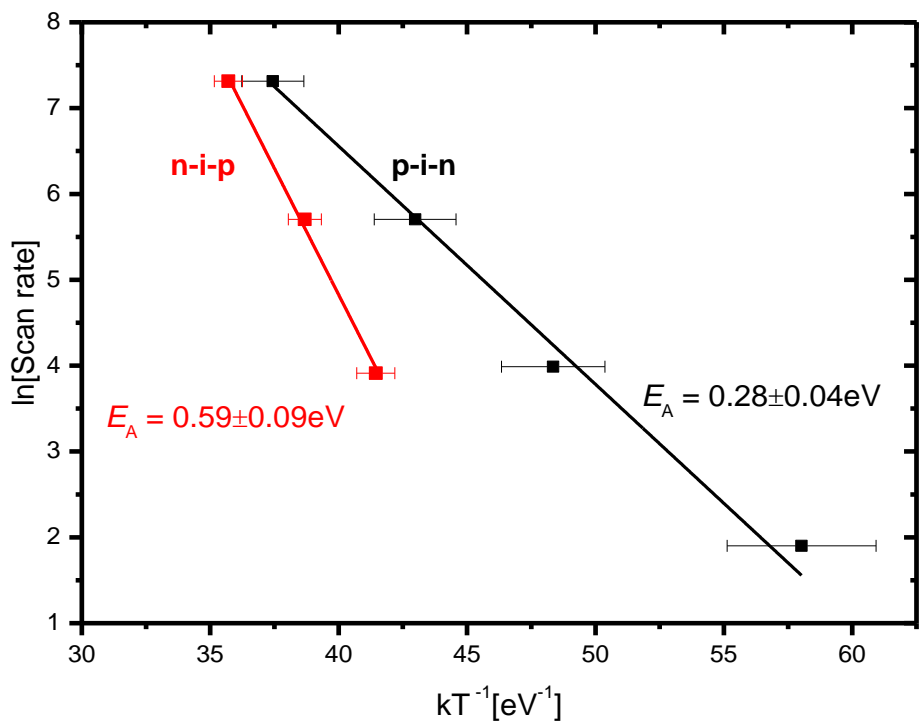

*Figure 9: Semi-log plot of scan rate vs. (kT)$^{-1}$ and the corresponding experimentally derived activation energies, $E_A$.*

Since $R$ (cf. Fig. 8) is an exponential function of $E_A$, $R = \frac{scan\ rate}{e^{-\frac{E_A}{kT}}}$, even slight changes of $E_A$ will induce order-of-magnitude changes in the values of $R$. If $R$ is indeed the variable that will solely determine the extent of the observed hysteresis, all HF values obtained for the different scan rates should fall on the same curve, which should have a peak shape (i.e. Gaussian), as shown in Figure 8. Thus, another method for extracting $E_A$ will be to find the value of $E_A$ for the p-i-n and n-i-p devices that will yield the best fit to a Gaussian curve, having all scan rates on the same plot. We find that by plotting HF as a function of log(R) using the obtained activation energies from Figure 9 can indeed yield a Gaussian shaped curve, as shown in Figure 10. The reason for the relatively low HF values in the 300 mV s$^{-1}$ scan rate close to the maximum for the n-i-p cell remains unclear, and this discrepancy might be resolved using more sophisticated definition of the HF, beyond the scope of the present analysis. Using other values for the activation energy results in poor fitting of the HF(R) curves to a Gaussian, as shown in Figure S9.

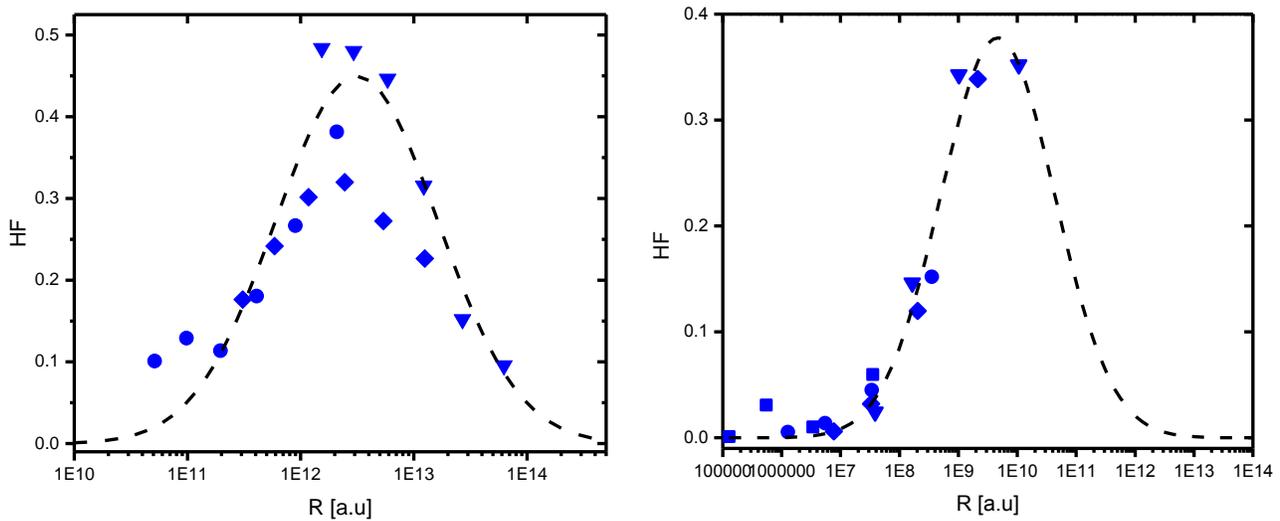

*Figure 10: Semi-log plot of HF as function R (cf. Fig. 8) for the n-i-p (a) and p-i-n devices (b), showing the different scan rates: 5 mV s$^{-1}$ (squares), 50 mV s$^{-1}$ (circles), 300 mV s$^{-1}$ (diamonds) and 1500 mV s$^{-1}$ (triangles).*

The main difference between the n-i-p and p-i-n devices is visualized nicely in Figure 10: for the n-i-p case, reaching low HF values could be obtained experimentally by using R values either far below the peak, i.e. $R = 10^{10}$ (low scan rate and high T) or far above the peak, $R = 10^{14}$ (high scan rate and low *T*), while for the p-i-n case, only the *R* values below the peak, positioned at $R = 5*10^9$, were observed experimentally. Furthermore, Figure 10 allows us to predict important scan parameters, i.e., for the n-i-p device, the scan rate in which a nearly hysteresis-free J-V curve should be observed at RT, using the value of $R = 10^9$ yields a scan rate of ~1 mV s$^{-1}$, slow enough for the slow process to follow completely the scan rate. Using the other extreme value of $R = 10^{14}$ yields ~100 V s$^{-1}$, fast enough so that the slow process will not cause any significant changes to the system during the scan and is also in good agreement with the result observed by Tress et al.[17]. Alternatively, using a typical scan rate of 50 mV s$^{-1}$, the temperature at which a hysteresis free *J-V* curve should be observed will be either above 373 K for $R = 10^9$, a temperature which is high enough for accelerating the slow process enough to follow the scan rate, or below 220 K for $R = 10^{14}$, decelerating the slow process so that no significant change will occur in the system during the scan. For the p-i-n device, Figure 10b allows to predict from which temperature a hysteresis-free J-V curve will be observed above the peak (located at $R = 5*10^9$). Taking a typical scan rate of 50 mV s$^{-1}$ and using the right extreme value of $R = 10^{12}$ yields ~140 K, or 160 K for the highest scan rate used in this study (1500 mV s$^{-1}$), 40 K below the lowest temperature at which the p-i-n device was measured (where maximum hysteresis was observed). Furthermore, using the observed peak position, the scan rate at which hysteresis will be at its maximum at RT can be calculated for the p-i-n cells. Using $R = 5*10^9$ and T=300 K yields a scan rate of ~$10^5$ mV s$^{-1}$.

There are three possibilities to describe the slow process, to which the above-calculated activation energy is attributed, in the p-i-n or n-i-p cell:

(1) drift / diffusion of ionic species <u>*within the perovskite layer*</u>;

(2) de-trapping of charge carriers from trap states close to the conduction band maximum (CBM) at the interface with the n-type layer and/or close to the valence band minimum VBM at the interface with the p-type layer;

(3) drift / diffusion of ionic species from the perovskite layer across/into the n or p interfaces.

While there is still much controversy regarding the physical process, to which the observed activation energy, $E_A$, should be attributed, our study shows that $E_A$ for the slow process can vary greatly even in the same laboratory under the same measurement conditions for different device architectures. $E_A$ values between 0.1 to 0.7 eV have been reported in the literature, as can be seen in table 1.

**TABLE 1:** Summary of reported values of the activation energy, $E_A$, for the slow process in MAPbI$_3$-based device configurations

| $E_A$ [eV] | Structure | Method | Attributed to | Reference |
|---|---|---|---|---|
| 0.36 | Lateral device | Photothermal-induced resonance (PTIR) microscopy | Electromigration of MA ions | 35 |
| 0.39-0.43 | Large pellets | Electrical conductivity | Ionic conductivity | 14, 36 |
| 0.50±0.02 | p-i-n cell | Thermally simulated current [TSC] | Deep traps in the perovskite film | 37 |

| 0.24, 0.66 | n-i-p cell | Temperature dependent Capacitance-frequency measurements | Trap states below the CBM in the perovskite film | 38 |
| --- | --- | --- | --- | --- |
| 0.12 | p-i-n cell | Chronoamperometry | Not mentioned | 21 |
| 0.60 | n-i-p cell | Chronoamperometry | Iodide vacancy migration | 13 |
| 0.42±0.09 | n-i-p cell | Chronoamperometry | Iodide defect migration | 39 |
| 0.28 | - | Calculation | Migration of MA$^+$ ions | 40 |
| 0.29 | - | Calculation | Migration of H$^+$ ions | 41 |

Both values of the activation energies in this study fall within the reported range of the values listed in table 1. Interestingly, the measured value of $E_A$ = 0.28±0.04 eV for the p-i-n devices is in excellent agreement with the calculated values of 0.28 eV for the defect migration of MA$^+$ ions[40] or 0.29 eV for H$^+$ migration,[41] and is also quite close to the experimentally observed value of 0.36 eV[35] attributed to MA$^+$ defect migration, derived from photothermal-induced resonance (PTIR) microscopy. The fact that the activation energy in the n-i-p device is higher than in the p-i-n device also agrees with the reported values, measured using chronoamperometry [13,21,39]. Relating the activation energy to option (1) above implies that the crystallization of the MAPbI$_3$ layer on TiO$_2$ and PEDOT:PSS is remarkably different, leading to such different morphologies that will affect the energy required for ion migration within the MAPbI$_3$ layer. This is quite unlikely, and was also ruled out recently by Kim et al., who observed that the hysteresis in the J-V curves does not correlate with the morphology of the perovskite film deposited on different substrates[12]. Option (2) implies that the measured activation energy is related to the de-trapping rate of charges at one of the interfaces. We exclude this option due to the fact that if the trap energy is indeed ~0.6 eV below the CBM for the n-i-p case, our simulations (Figure 7c) show that in the studied temperature range (280-330K) such deep traps will be nearly full, and the de-trapping rate will be negligible.

Option (3) requires different rates of permeabilities across (if they are permeable at all) and/or accumulation of ions at one or both of the interfaces. These rates should depend on the different permeabilities/reactivities of the n or p contact layers to/with the migrating ions. In the case of the p-i-n device, both the n- and p-type contact layers are comprised of non- or at best semi-crystalline, low density organic materials, PCBM and PEDOT:PSS, compared to the n-i-p device in which only the p-type layer, spiro-OMETAD, is such an organic, while the n-type layer is the denser inorganic, $TiO_2$. Since it is known that organic semiconductors can be easily doped using even large ions, It is possible that $MA^+$ or $I^-$ ions from the perovskite layer can diffuse/migrate into the organic layers (PCBM or PEDOT:PSS) depending on the permeability of that layer to the specific ion, as recently suggested by De Bastiani et al.[42] This possibility, which awaits actual experimental proof, in turn, can limit accumulation of excess ions at the interfaces and any resulting trap formation due to this excess, thus limiting interfacial recombination. However, in the n-i-p case, ions might accumulate near the $TiO_2$ interface due to the expected lower permeability of $TiO_2$ compared to organic contact materials. Such accumulation would change the electrostatic potential profiles at the interfaces and affect trap formation, leading to hysteresis, which is already observed at RT for devices containing $TiO_2$[10,11]. At lower temperatures the permeability of the organic layers in the p-i-n device to ionic species will probably be lower than at RT, leading to more ion accumulation as the temperature is lowered, resulting in trap formation and hysteresis at low temperature. It is important to note that the temperature dependence observed in the simulations can still be present in the experiments, but, if the experimental results are dominated by ion migration effects then the temperature dependence of detrapping cannot be measured separately by our experiments, and is a subject of future study.

To estimate the rate of diffusion of the migrating ion in the n-i-p or p-i-n cells, we use the 1-D diffusion approximation: $L = \sqrt{D\tau}$, where D is the diffusion coefficient, L the diffusion length and $\tau$ the diffusion time, which we will now estimate and compare to the dwell time at each step during the J-V scan. We then use $\tau^{-1}$ to get a rate, which we take as the rate that determines the system response. We consider the following two scenarios: (1) the diffusing ion moves throughout the entire perovskite layer, i.e. *L* ~ 300 nm (typical absorber layer width in solar cells, made with $MAPbI_3$); (2) the diffusing ion moves across the interface, i.e., *L* ~ 3 nm, while using D values that were deduced/estimated in the literature, which are ~$10^{-8}$ [14] and ~$10^{-12}$ [13] cm² s⁻¹; the resulting $\tau$ for each of the cases is shown in the table below:

**TABLE 2:** Estimated system response rates based on 1-D diffusion approximation

|  | 1 | 2 | 3 | 4 |
|---|---|---|---|---|
| D [cm² s⁻¹] | $10^{-8}$ | $10^{-8}$ | $10^{-12}$ | $10^{-12}$ |
| L [nm] | 300 | 3 | 300 | 3 |
| $\tau$ [sec] | 0.09 | 0.000009 | 900 | 0.09 |
| $\tau^{-1}$ [s⁻¹] | ~ 10 | ~ $10^5$ | ~ $10^{-3}$ | ~ 10 |

For the n-i-p cells, in the J-V scan, the relevant timescale to observe hysteresis in the studied scan rates is between 0.01-0.2 sec (calculated by dividing the voltage step (10mV) by the scan rate (1500mV s⁻¹ or 50mV s⁻¹); based on table 2, option (1) –diffusion through the entire perovskite layer with D~$10^{-8}$ cm² s⁻¹, or option (4) – diffusion only through the interface with D~$10^{-12}$ cm² s⁻¹, are the only relevant cases for which hysteresis will be observed in the J-V scans. However, option (1) is excluded, because if this were the case the activation energies for the n-i-p and p-i-n cells should be very similar to identical. This leaves option (4), i.e., the rate-determining step is diffusion across the interface, with D~$10^{-12}$ cm² s⁻¹ for the n-i-p cells, supporting our earlier conclusion of the critical role of the interfaces. This value agrees with the one that was calculated by Eames et al.[13], but is different from that deduced by Yang et al.[14] The latter used large pellets between carbon contacts, i.e., without any –n or –p contact layers.

For the p-i-n cells, the relevant timescale to observe hysteresis at RT can be estimated using the calculated scan rate from Figure 10b, which was found to be $10^5$ mV s⁻¹, leading to a timescale of ~0.1msec. Using this value to calculate the diffusion coefficient yields D~$10^{-9}$ cm² s⁻¹ for these cells, 3 orders of magnitudes higher than that estimated for the n-i-p cells. Interestingly, this value is quite close to that obtained by Yang et al.[14] (2.4*$10^{-8}$ cm² s⁻¹), from experiments with Carbon-based electrodes. Such electrodes may have permeability for the migrating ions from the halide perovskite, roughly comparable to the organic PCBM and/or PEDOT-PSS contact layers that we used in the p-i-n cell.

One of the most striking differences between the observed hysteresis in the p-i-n devices compared to the n-i-p ones is the lack of hysteresis in the Voc in the p-i-n devices. Our simulation results in Figure 7b show that hysteresis in Voc is only observed for relatively strong hysteresis effects, i.e. for trap depths > 0.3 eV. Comparing Figure 7b with the experimental results suggests relatively deep trap states are present in the n-i-p cell as opposed to the p-i-n one, leading to much stronger hysteresis in general in the n-i-p cells.

The fact that for the p-i-n cells, $V_{OC}$ shows very little dependence on scan direction or scan rate due to the very small hysteresis observed in this parameter, allows us to analyze the temperature dependence of the $V_{OC}$ of a fresh p-i-n cell down to 80 K in a systematic way. Such analysis can provide details on the recombination mechanism that occurs in the solar cell, and has been used previously to this end for CdTe, CIGS, a-Si and other types of thin film solar cells[43,44,45]. As a reference point we used a "state of the art" GaAs solar cell (28% eff., NREL) and its $V_{OC}$ dependence on T is shown in the inset of Figure 6:

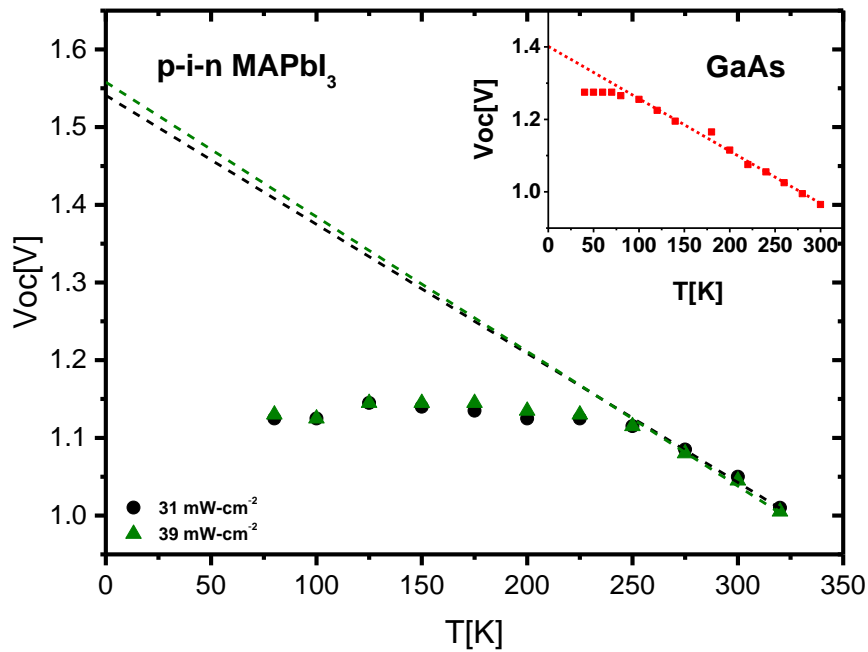

*Figure 11: $V_{OC}$ as a function of temperature for a fresh p-i-n cell at 31 mW cm$^{-2}$ (black circles) and 39 mW cm$^{-2}$ (blue triangles). Inset: $V_{OC}$ as a function of temperature for a state of the art (28% efficient) GaAs cell (red squares, at 60 mW cm$^{-2}$ illumination intensity).*

The temperature dependence of $V_{OC}$ shows the expected increase of the $V_{OC}$ as the temperature is lowered[45] with a temperature coefficient of close to 2 mV K$^{-1}$, a typical value for photovoltaic cells (see Table 2 below). When approaching 200 K, the $V_{OC}$ slightly deviates from the trend, and seems to saturate at a maximum value around 1.15 V; this saturation will be discussed below.

In the ideal case (neglecting series and shunt resistance effects), when extrapolating to 0 K, the $V_{OC}$ should reach an $E_G/e$ value which, at room temperature, is ~1.6 V for MAPbI$_3$[3] (the band gap decreases only slightly upon cooling through the phase transition from tetragonal to orthorhombic phase[46]), based on Equation (2):[47]

$$V_{OC} = \frac{E_g(T)}{e} - \frac{nkT}{e}\left(\ln\left[\frac{J_{s0}}{J_F}\right]\right) \qquad (2)$$

where $n$ is the diode ideality factor, $J_{S0}$ is the reverse saturation current and $J_F$ is the photocurrent. As can be seen from the inset of Figure 11, the temperature dependence of the studied GaAs cell obeys Equation (2), and by extrapolating the $V_{OC}$ to $T$=0 K, we get $V_{OC}$ = 1.40 V in good agreement with the (RT) bandgap of GaAs which is 1.42 eV, suggesting very minor interfacial recombination.[43] This result implies that the dominant recombination mechanism occurs in the space charge region of the absorber layer,[45,48] as expected for such an efficient GaAs solar cell (we also note that even at T=80 K, when performing the J-V scan at the highest scan rate tested of 1.5 V sec$^{-1}$, no hysteresis was detected for the GaAs cell). If we repeat the same exercise for the MAPbI$_3$ p-i-n cell, while ignoring the points below 225 K (discussed later), we obtain a value of 1.55 ± 0.01 eV, which agrees well with the experimental band gap and with a very recent result obtained for a mesoscopic MAPbI$_3$ cell (which showed hardly any hysteresis) by Leong et al.[49], and resembles the result of the GaAs cell, also suggesting that the dominant recombination mechanism occurs in the space charge region of the absorber layer. In our case this could be the entire MAPbI$_3$ layer since it has been shown that for good MAPbI$_3$ cells the MAPbI$_3$ is nearly intrinsic.[4] Such low interfacial recombination suggests very low trap density at the interfaces, or the existence of very shallow traps in the case of the p-i-n device. This might explain why hysteresis in these devices does not appear at RT, but, rather, at low T, where detrapping of charges at the interface is less efficient, and why there is little if any hysteresis in the $V_{OC}$ for the p-i-n devices in this study. However, we also note that if all recombination was absent at the heterojunctions, as well as within the absorber space charge region (SRH recombination), we would expect the luminescence efficiency of these devices to be very high, and the open-circuit voltage to approach 1.25 V at room temperature.[50] Since this is not the case[51], we expect there still to be recombination at the heterojunctions, but not via deep trap states.

The observed saturation of the $V_{OC}$ at 200 K can be rationalized as follows: Even in the absence of a significant density of trap states, the system can reach the maximum splitting of the quasi-Fermi levels upon illumination (i.e., the $V_{OC}$ limit), because the built-in potential of the device will be dictated by the band gap, doping density and Fermi level positions at the contacts of the $MAPbI_3$. When this limit is reached, the system will not anymore follow Equation (2). This can be confirmed by analyzing the saturation of the Voc as a function of illumination intensity and temperature (Fig S7), and it can be seen that at higher light intensities, the saturation temperature is shifted to higher values without changing the saturation value of the Voc, thus confirming that we have reached the maximum quasi-Fermi level splitting which is limited by the contact materials.

Another important parameter which can be extracted from Figure 11, when a well-behaved temperature dependence of the $V_{OC}$ is observed as is the case here, is the temperature coefficient of the cells, which is determined from the slope (averaged over 3 different p-i-n cells), and is found to be -1.7±0.4 mV $K^{-1}$, and is compared to other types of solar cells in Table 2:

**TABLE 2:** Dependence of $V_{OC}$ on temperature for several types of solar cells

| Cell type | Temperature coefficient [mV K$^{-1}$] | Reference |
|---|---|---|
| **CIGS** | −2.0 to −3.3 | 52 |
| **Crystalline Si p-n junction** | -1.9 to -2.5 | 53 |
| **CdTe** | -2.1 to −2.2 | 54 |
| **Solid-state dye-sensitized TiO$_2$:Spi-roMeOTAD** | -1.4 | 55 |
| **Bulk-heterojunction organic P3HT:PCBM** | -0.6 to -0.7 | 56 |
| **p-i-n MAPbI$_3$** | -1.3 to -2.1 | This study |
| **GaAs** | -1.6 to -1.7 | This study |

Although the p and n contact layers in the MAPbI$_3$ cells in this study are materials which are used for organic solar cells (PEDOT:PSS and PCBM), it can be seen that the temperature coefficient of the p-i-n perovskite solar cells is much larger than that of organic BHJ (bulk-heterojunction) cells (-0.6 to -0.7mV K$^{-1}$), and similar to that of other inorganic solar cells such as GaAs, CIGS, Si and CdTe cells, suggesting that the working mechanism of the p-i-n devices is similar in nature to that of other thin film crystalline solar cells.

**Conclusions**

In summary, by performing a detailed study on the effect of the scan rate, temperature and light intensity on the J-V characteristics of MAPbI$_3$-based solar cells, we were able to conclude that hysteresis *can* be observed for p-i-n, PEDOT:PSS / MAPbI$_3$ / PCBM devices under specific working conditions. We deduce that ion migration at the interfaces of the perovskite layer is the rate-determining step in both n-i-p and p-i-n devices, and the rate of such migration is highly dependent on the contacting materials, thus leading to <u>*different*</u> activation energies between the two types of solar cells. We also stress that the main differences between the n-i-p and the p-i-n devices originate from the different contact materials, organic non- or semicrystalline versus more crystalline and denser inorganic oxide, rather than the configuration of the device. We conclude that controlling the interfaces using different -p or -n contact materials can suppress the observable hysteresis at a certain scan rate or temperature, as many studies have shown. However, the main cause for hysteresis will remain due to the presence of mobile ions within the MAPBI$_3$ layer. Thus any device that contains MAPbI$_3$ is prone to hysteresis in the J-V curve, and the contacting materials will only determine at which scan parameters the hysteresis in the J-V curve will be observed experimentally.

Our results can be understood by connecting the emergence of hysteresis to the different permeabilities to mobile ionic species in the MAPbI$_3$ into the n and p contact layers of the devices, with different ionic diffusion coefficients that depend on the contact material, an idea that requires further research, yet already explains the wide spread of reported activation energies and diffusion coefficients for MAPbI$_3$-based devices. In any cases, our findings imply that the combination of PCBM and PEDOT:PSS contact layers generates superior electronic interfaces compared to that of TiO$_2$ and spiro ones, suggesting less limitations on the $V_{OC}$ in such devices, and the potential to induce a larger built-in field which might also improve other types of cells such as ones based on CsPbBr$_3$ and MAPbBr$_3$.

*Submitted to Advanced Energy Materials, November 1st, 2015*

## Supplementary information

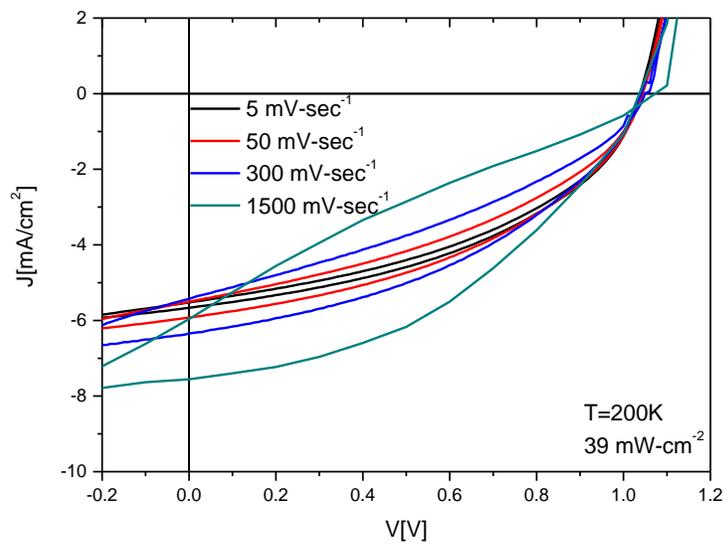

*Figure S1: J-V curves measured for the p-i-n cell under 39 mW cm$^{-2}$ light intensity at different scan rates at 200K*

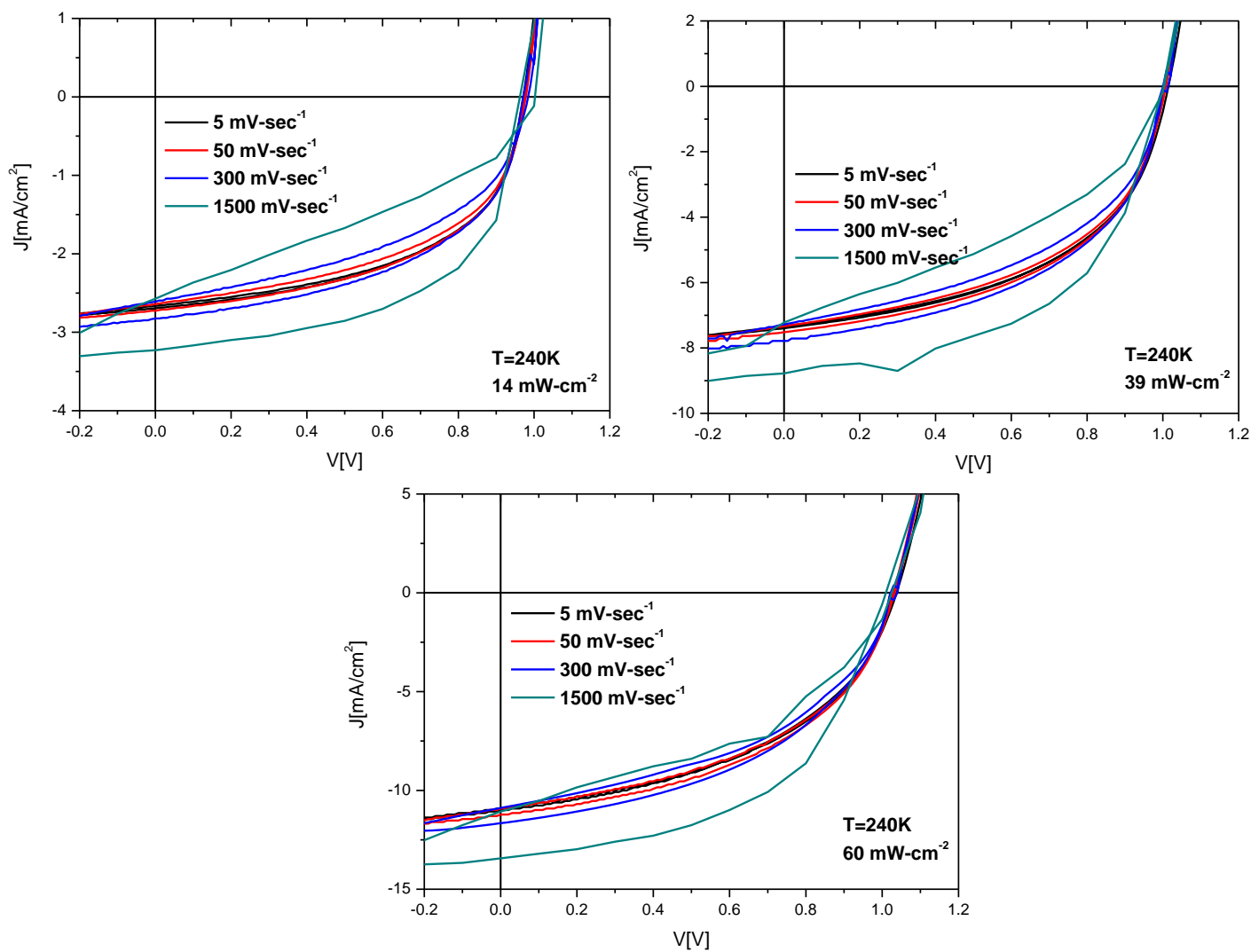

*Figure S2: J-V curves for the p-i-n cell measured at 240K under 14 mW cm$^{-2}$ (top left), 39 mW cm$^{-2}$ (top right) and 60 mW cm$^{-2}$ (bottom)*

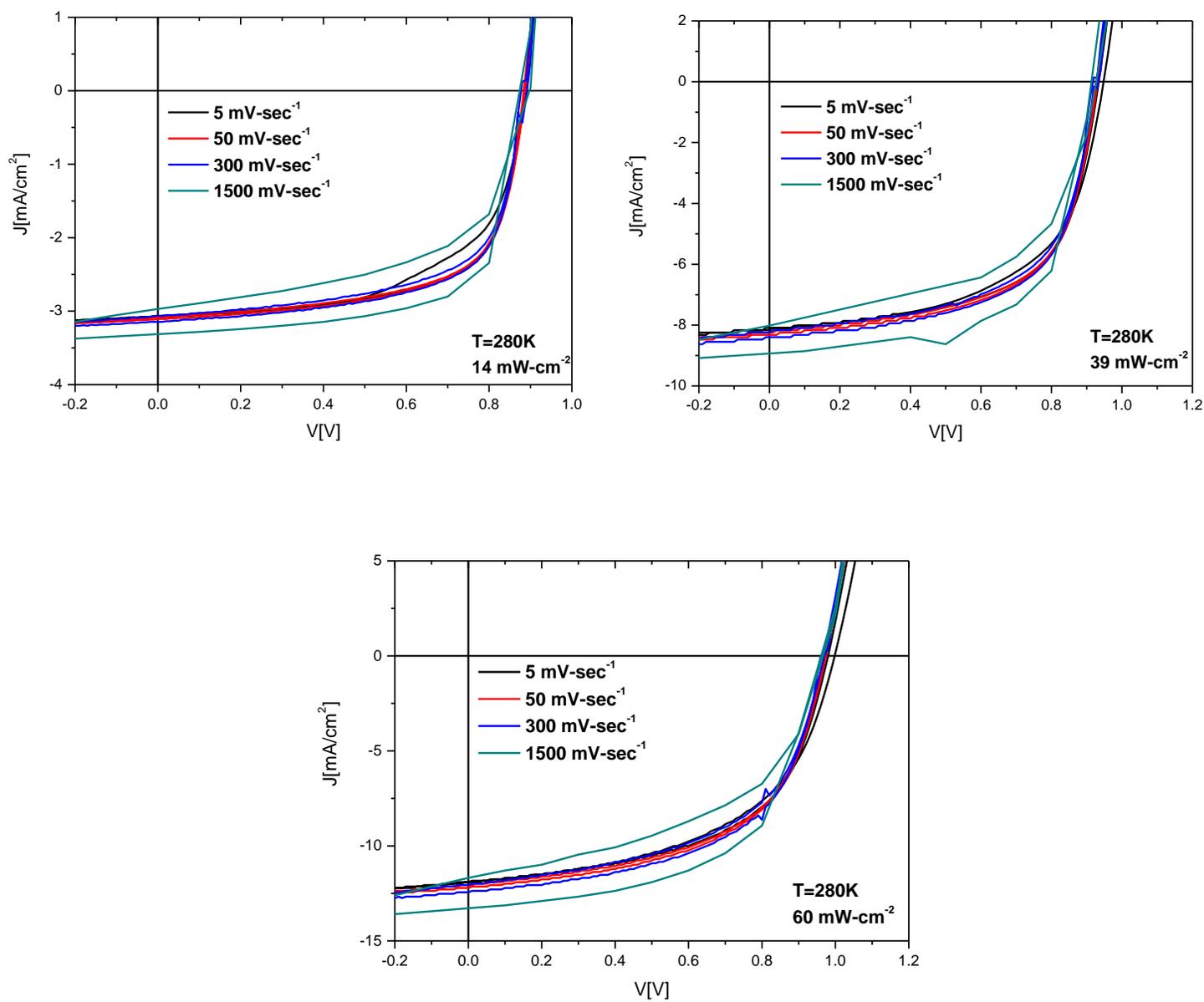

*Figure S3: J-V curves for the p-i-n cell measured at 280K under 14 mW cm$^{-2}$ (top left), 39 mW cm$^{-2}$ (top right) and 60 mW cm$^{-2}$ (bottom)*

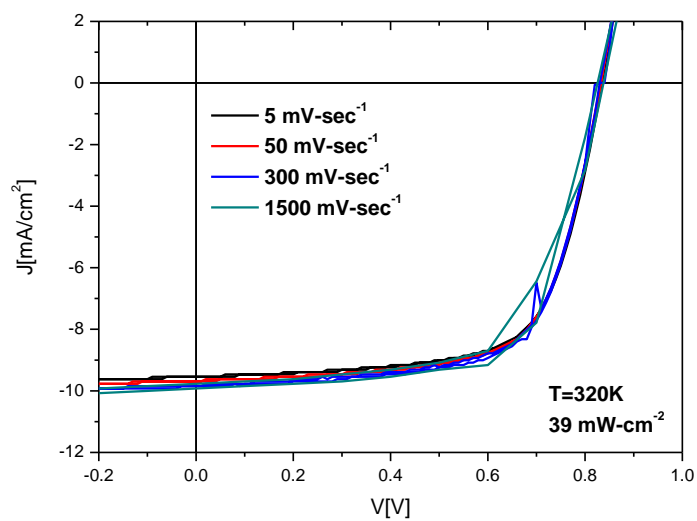

*Figure S4: J-V curves for the p-i-n cell measured under 39 mW cm$^{-2}$ light intensity at different scan rates at 320K*

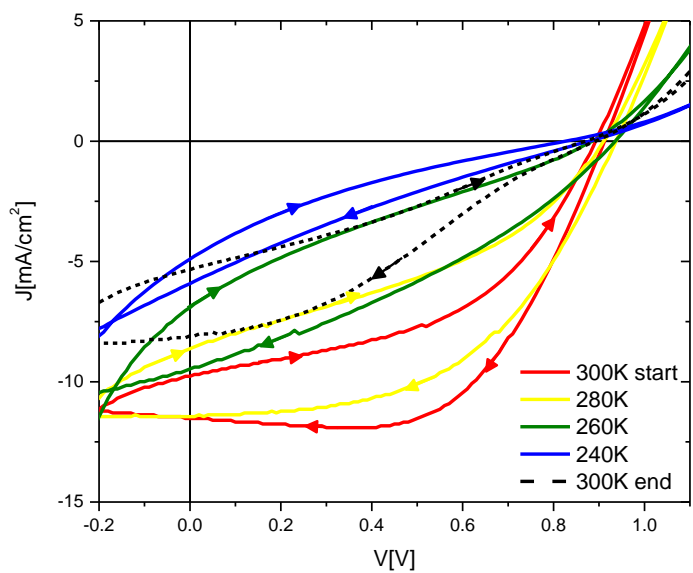

*Figure S5: J-V curves of n-i-p device, measured under light intensity of 60 mW cm$^{-2}$ at a scan rate of 50 mV sec$^{-1}$. The dashed line represents the J-V curve taken after heating back up to 300K; arrows indicate the scan directions.*

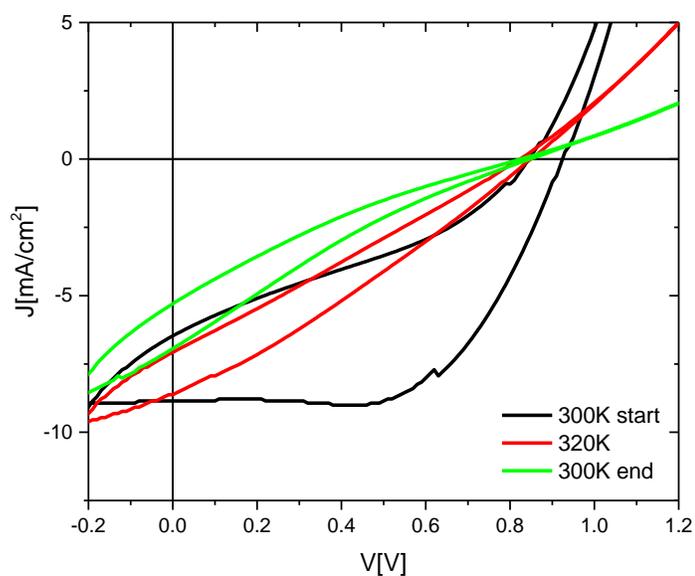

*Figure S6: J-V curves of n-i-p device measured under light intensity of 60 mW cm$^{-2}$ at a scan rate of 50 mV sec$^{-1}$ under vacuum.*

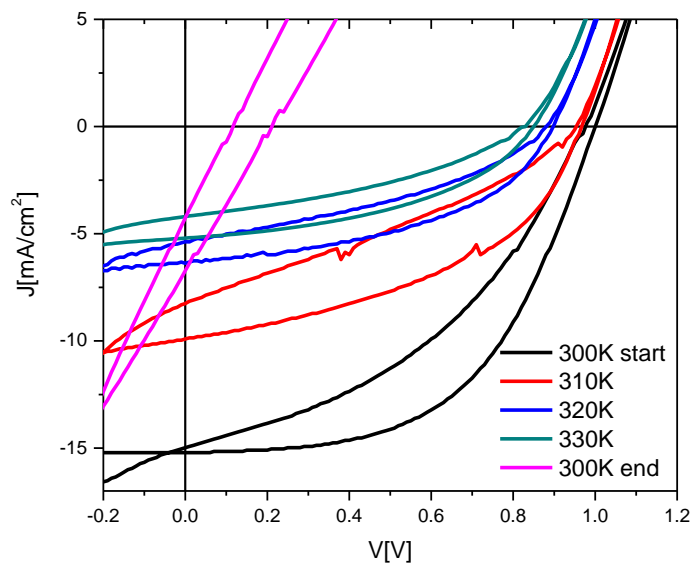

*Figure S7: J-V curves of n-i-p device, which underwent heating under ambient conditions, measured under light intensity of 60 mW cm$^{-2}$ at a scan rate of 50 mV sec$^{-1}$*

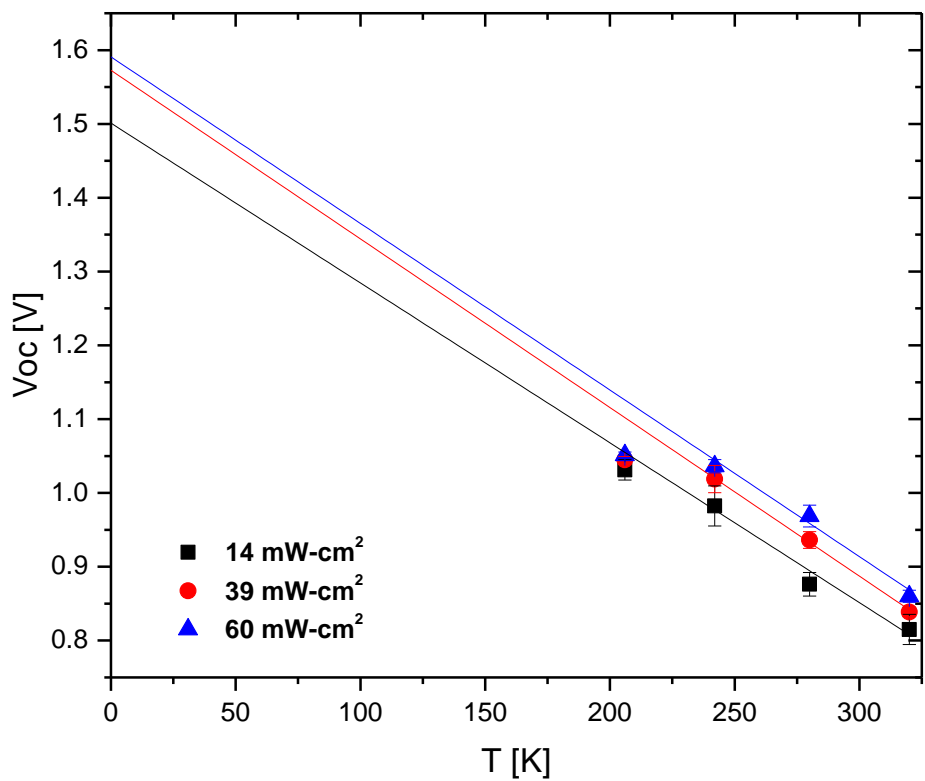

*Fig. S7: Voc as a function of temperature for the p-i-n device at the 3 different light intensities that were used to extract the HF values. The values in each temperature are an average of the forward and reverse scans (for the four different scan rates). The error bars denote the standard deviation of the different values.*

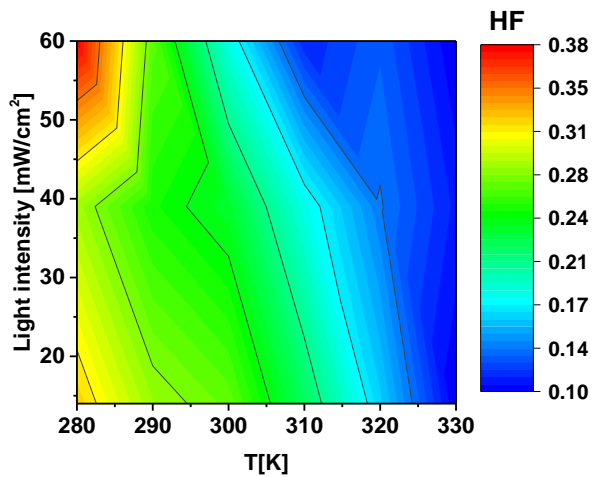

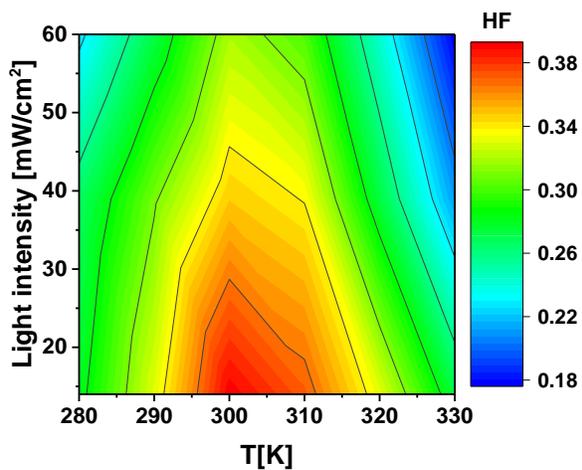

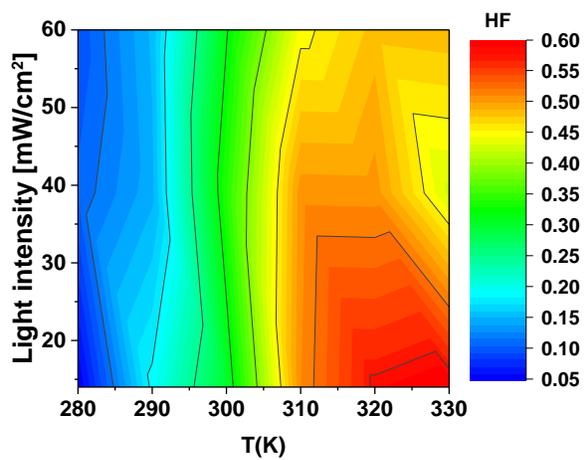

*Fig. S8: Hysteresis factor for n-i-p devices as a function of light intensity and temperature for scan rates of 50 (a) 300 (b) and 1500 mV s$^{-1}$ (c). The color map indicates the value of the hysteresis factor **with a separate scale for each of the scan rate**. While this prevents direct comparison between the different scan rates (which is possible for the figure in the main text) it provides more detail for each scan rate.*

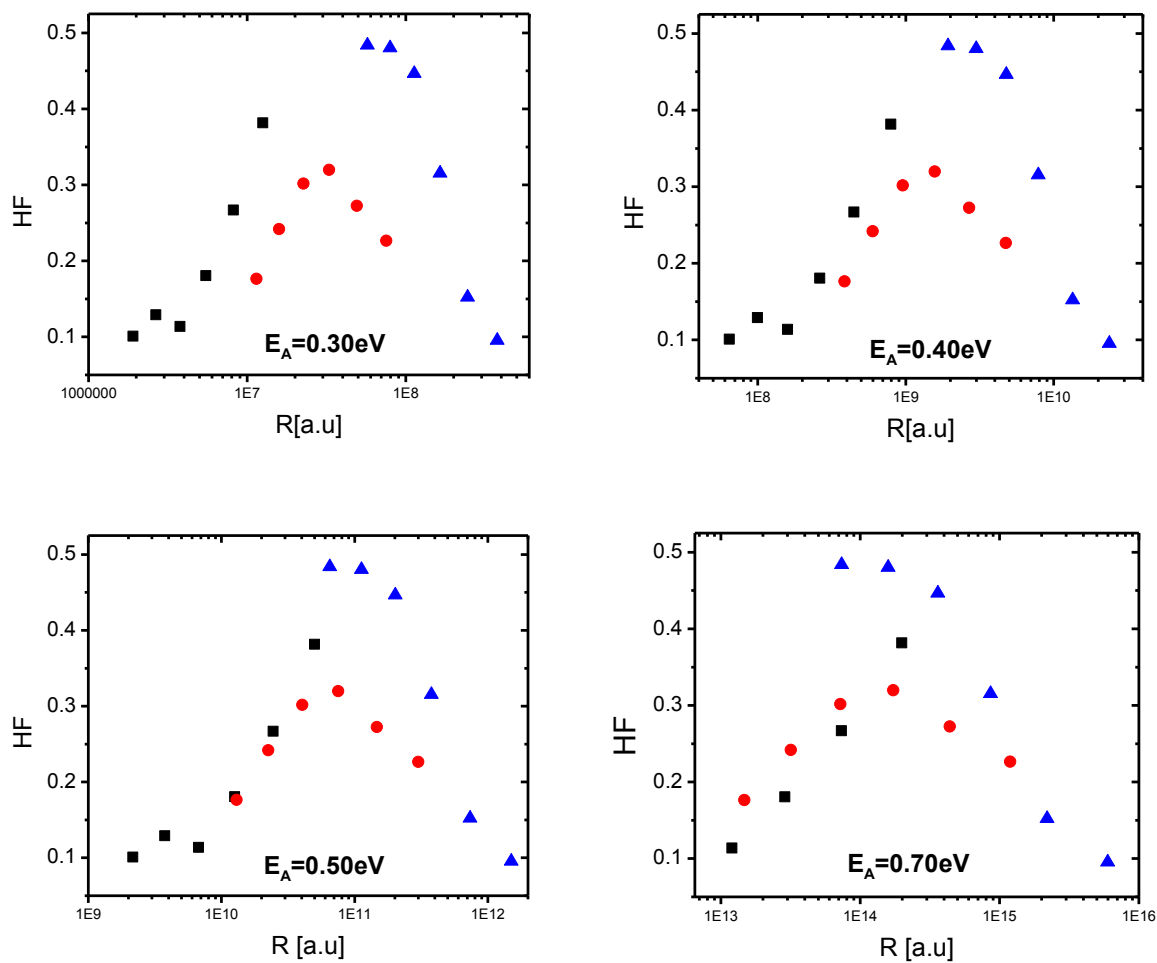

*Fig. S9: HF as a function R using different $E_A$ values for the n-i-p device,*

# Modelling traps in perovskite solar cells

A. The numerical model to describe perovskite solar cell operation

To model the device operation of perovskite solar cells a numerical model is used which describes carrier injection, transport, generation, recombination, and trapping for electronic and ionic charges.[1-4] This model solves the Boltzmann transport equations and Poisson's equation on a 2D grid by forward integration in time. In the next paragraphs a detailed description is given on the methods to describe charge transport, carrier injection and carrier recombination. Additional optional processes like i) charge trapping and ii) trap-assisted recombination are included in the numerical model as well. Parameters that need to be defined for each calculation are shown in Table A1. Here CB and CV are the conduction band and valence band, respectively.

TABLE A1. Input parameters for the numerical model of a perovskite solar cell.

| Parameter | Symbol |
| --- | --- |
| Mobility (electron; hole; anion; cation) | $\mu_n$ ; $\mu_p$ ; $\mu_a$ ; $\mu_c$ |
| CB and CV of semiconductors | $E_{HOMO}$ ; $E_{LUMO}$ |
| Density of states of semiconductors | $N_0$ |
| Anion and cation density | $a$ ; $c$ |
| Work function of contacts | $\varphi$ |
| Bias voltage | $V_{bias}$ |
| Temperature | $T$ |
| Relative dielectric constant | $\varepsilon_r$ |

A.1 Charge transport

In the 2D model a rectangular grid is used with grid points at the corners of each cell. The areas of the cells are determined by the widths of the columns and rows which can be specified per row or column. For each cell a material type is specified, either gate, contact, dielectric or semiconductor. Calculations of current, recombination, etc. are performed on the grid points. The 4 cells surrounding the grid point are in equilibrium with each other, which is effectuated by area-average weighting of these 4 cells. Consequently, a single quasi Fermi level is defined on each grid point.

Carrier transport is described by the following Boltzmann transport equations:

$$J_i = -q\mu_i n_i \nabla \psi_{F,i}, \qquad \text{(S1)}$$

where $q$ is the elementary charge, $\mu_i$ and $n_i$ are the mobility and density for free electrons ($n_i = n$), free holes ($n_i = p$), free anions ($n_i = a$), or free cations ($n_i = c$) and $\psi_{F,i}$ is the quasi Fermi energy which is given by

$$\psi_{F,n} = \psi_L - \frac{kT}{q}\ln\frac{n}{N_0}, \qquad \text{(S2)}$$

$$\psi_{F,p} = \psi_H + \frac{kT}{q}\ln\frac{p}{N_0}, \qquad \text{(S3)}$$

$$\psi_{F,a} = V - \frac{kT}{q} \ln \frac{a}{N_0}, \tag{S4}$$

$$\psi_{F,c} = V + \frac{kT}{q} \ln \frac{c}{N_0}, \tag{S5}$$

where $\psi_L = E_{CB} + V$ and $\psi_H = E_{VB} + V$ and *V* is the electrostatic potential. *k* is Boltzmann's constant and *T* is the temperature. Einstein's relation between the diffusion constant and the mobility is assumed to hold. The ion current is zero at the electrodes, for the electrodes are ionically blocking. The continuity equations are

$$\frac{dn}{dt} = \frac{1}{q} \frac{dJ_n}{dx} - R, \tag{S6}$$

$$\frac{dp}{dt} = -\frac{1}{q} \frac{dJ_p}{dx} - R, \tag{S7}$$

$$\frac{da}{dt} = \frac{1}{q} \frac{dJ_a}{dx}, \tag{S8}$$

$$\frac{dc}{dt} = -\frac{1}{q} \frac{dJ_c}{dx}, \tag{S9}$$

for electrons, holes, anions, and cations, respectively. Here $J_i$ is the current density of particle *i* and *R* is the recombination rate. The potential profile throughout the device is calculated from Poisson's equation:

$$\nabla^2 V = -\frac{q}{\varepsilon_0 \varepsilon_r}(p - n + c - a), \tag{S10}$$

where $\varepsilon_0$ and $\varepsilon_r$ are the dielectric and the relative dielectric constant, respectively.

### A.2 Carrier injection

To describe the injection of electrons and holes into the semiconductors, the following parameters are of importance: the VB and CB levels of the semiconductor; the Fermi level of the contacts; and the charge carrier density in the semiconductor at the interface. The injection model is based on Boltzmann injection:

$$n(1) = N_0 \exp\left(-\frac{q\varphi_n}{kT}\right), \tag{S11}$$

where $n(1)$ is the electron density at the first grid-point next to the electrode and $\varphi_n$ is the injection barrier height. In this model injection is not affected by the grid-point spacing.

### A.3 Charge trapping

The charge trapping rate for electrons ($T_n$) is described by the product of the free electron density (*n*), the available trap density ($N_t$-$n_t$), and a trapping coefficient (in m$^3$s$^{-1}$):

$$T_n = c_n \cdot (N_t - n_t) \cdot n. \tag{S12}$$

Detrapping of electrons ($D_n$) is described by:

$$D_n = c_n \cdot N_0 \cdot n_t \cdot exp\left(-\frac{E_t}{kT}\right), \tag{S13}$$

where $N_0$ is the density of states and $E_t$ is the trap energy. Empty traps are considered neutral.

### A.4 Recombination

Recombination of free electrons and free holes can be described by a Langevin process, including a (Langevin) prefactor as described in ref [5]:

$$R_f = L_{pre}\frac{q(\mu_n+\mu_p)}{\varepsilon_0\varepsilon_r} \cdot p \cdot n, \tag{S14}$$

where $R_f$ is the recombination rate of free carriers and $L_{pre}$ is the prefactor.[5]

Recombination between trapped electrons and free holes is also described by a Langevin process in the model:

$$R_t = L_{pre,t}\frac{q(\mu_p)}{\varepsilon_0\varepsilon_r} \cdot p \cdot n_t = c_n \cdot p \cdot n_t, \tag{S15}$$

where $R_t$ is the recombination rate of free holes and trapped electrons, $n_t$ is the trapped electron density, $c_n$ is the recombination coefficient for this recombination process, and $L_{pre,t}$ the Langevin prefactor for this specific recombination process.

### A.5 Used parameters

Here follow the parameters we used to model a perovskite solar cell. A grid was used with 4 nm grid-point spacing (see Fig. A1). Model parameters for the electron transport material (ETM), the perovskite, and the hole transport material (HTM) are shown in Table A2 and are based on $TiO_2$[6] and Spiro-OMeTAD, respectively. For the perovskite, an equal electron ($\mu_n$) and hole ($\mu_p$) mobility was chosen similar to literature values.[7,8] The dielectric constant $\varepsilon_r$ was set at 6.5.[9] A weakened form of Langevin recombination was assumed by addition of a prefactor $L_{pre} = 10^{-5}$ in the recombination rate equation.[5] Homogeneous generation $G$ of free electrons and holes was assumed throughout the perovskite layer 2.5 $nm^{-3}s^{-1}$ to model light conditions roughly equal to 1 sun.[10] Doping in the transport layers was modelled by a homogeneous density of immobile anions (a) or cations (c) which are electrostatically compensated by mobile electrons (n) and holes (p).

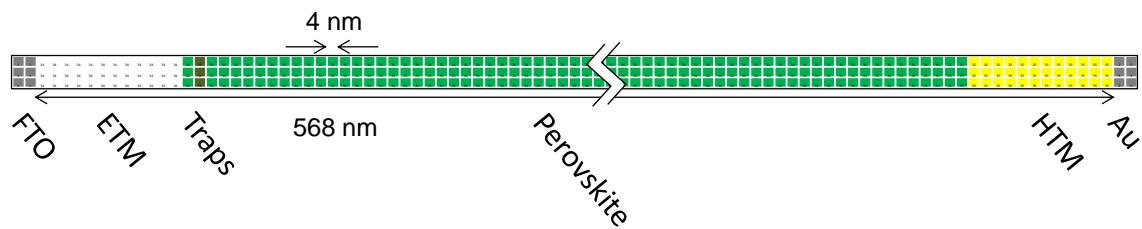

FIG. A1. Grid layout of the modelled perovskite solar cell.

The energy level diagram is shown in Fig. A2.

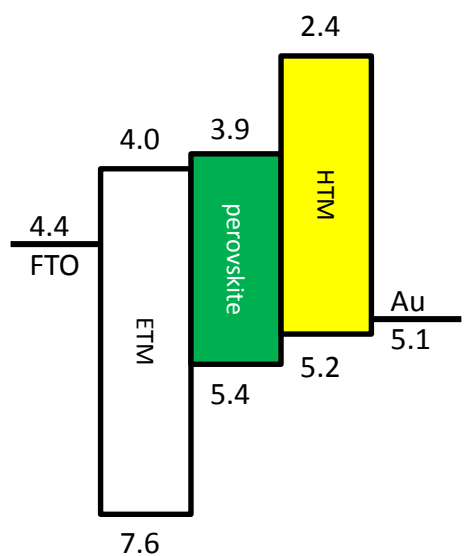

FIG. A2. Energy level diagram in eV of the simulated perovskite solar cell.

TABLE A2. General device model parameters.

| property | unit | ETM | Perovskite | HTM |
|---|---|---|---|---|
| $\mu_p, \mu_n$ | cm$^2$V$^{-1}$s$^{-1}$ | 0.02 | 2 | 0.02 |
| $\varepsilon_r$ | - | 80 | 6.5 | 3 |
| $L_{pre}$ | - | 1 | $10^{-5}$ | 1 |
| G | cm$^{-3}$s$^{-1}$ | 0 | $2.5 \cdot 10^{21}$ | 0 |
| initial $a$ | cm$^{-3}$ | 0 | X | $1 \cdot 10^{17}$ |
| initial $c$ | cm$^{-3}$ | $4 \cdot 10^{17}$ | X | 0 |
| Initial $n$ | cm$^{-3}$ | $4 \cdot 10^{17}$ | 0 | 0 |
| initial $p$ | cm$^{-3}$ | 0 | 0 | $1 \cdot 10^{17}$ |
| Layer thickness | nm | 48 | 472 | 48 |
| $N_0$ | cm$^{-3}$ | $10^{20}$ | $10^{20}$ | $10^{20}$ |

Trapping of electrons at the interface was described by equation S12. The trap energy was set at 0.2 eV at the n-type interface (see Fig. A1.). The trap density was set at $10^{17}$ cm$^{-3}$. The trapping coefficient $c_n$ was approximated by the product of the site volume (i.e. nDOS$^{-1}$ = 1/2.2·10$^{24}$ = 5·10$^{-25}$ m$^3$ following [7]) and the attempt frequency which can be related to the phonon frequency, which is typically 10$^{-10}$ – 10$^{-12}$ s$^{-1}$. Hence $c_n$ was set at 6x10$^{-13}$ m$^3$s$^{-1}$ for interface traps.

Recombination between trapped electrons and free holes in the bulk perovskite was described by equation S14 where $L_{pre,t}$ was maintained at 10$^{-5}$,[5] similar to recombination between free electrons and holes in the perovskite.

### A.6 Model operation

Initially no electrons or holes are present. Only at the electrodes electrons and holes are present due to the boundary conditions imposed by the injection model. In addition, the bias voltage is applied to one of the contacts (Au), whereas the other contact (FTO) is set at 0 V. For a defined number and size of time steps, the model then calculates the Fermi energy of all the carriers, solves the Boltzmann transport equations, accounts for electron-hole recombination, and solves the continuity equations to determine the new carrier densities at each grid point. A steady-state solution is obtained when the current, which is determined by the sum of current through the contacts, becomes constant.